\documentclass[twocolumn]{aastex631}

\newcommand\galight{{\sc galight}}
\newcommand\sersic{S\'ersic}
\newcommand\reff{R$_{\rm eff}$}
\newcommand\smass{M${_*}$}

\shorttitle{Decomposition study of $z>$6 quasar host galaxies}
\shortauthors{Ding et al.}
\graphicspath{{./}{figures/}}

\begin{document}

\title{SHELLQs-JWST Unveils the Host Galaxies of 12 Quasars at $z>6$}

\author[0000-0001-8917-2148]{Xuheng Ding}
\affiliation{School of Physics and Technology, Wuhan University, Wuhan 430072, China}
\affiliation{Kavli Institute for the Physics and Mathematics of the Universe (WPI), The University of Tokyo
Institutes for Advanced Study, The University of Tokyo, Kashiwa, Chiba 277-8583, Japan}

\author[0000-0003-2984-6803]{Masafusa Onoue}
\affiliation{Waseda Institute for Advanced Study (WIAS), Waseda University, 1-21-1, Nishi-Waseda, Shinjuku, Tokyo 169-0051, Japan; Center for Data Science, Waseda University, 1-6-1, Nishi-Waseda, Shinjuku, Tokyo 169-0051, Japan}
\affiliation{Kavli Institute for the Physics and Mathematics of the Universe (WPI), The University of Tokyo
Institutes for Advanced Study, The University of Tokyo, Kashiwa, Chiba 277-8583, Japan}

\author[0000-0002-0000-6977]{John D. Silverman}
\affiliation{Kavli Institute for the Physics and Mathematics of the Universe (WPI), The University of Tokyo
Institutes for Advanced Study, The University of Tokyo, Kashiwa, Chiba 277-8583, Japan}
\affiliation{Department of Astronomy, Graduate School of Science, The University of Tokyo, 7-3-1 Hongo, Bunkyo-ku, Tokyo 113-0033, Japan}
\affiliation{Center for Data-Driven Discovery, Kavli IPMU (WPI), UTIAS, The University of Tokyo, Kashiwa, Chiba 277-8583, Japan}
\affiliation{Center for Astrophysical Sciences, Department of Physics \& Astronomy, Johns Hopkins University, Baltimore, MD 21218, USA}

\author[0000-0001-5063-0340]{Yoshiki Matsuoka}
\affiliation{Research Center for Space and Cosmic Evolution, Ehime University, 2-5 Bunkyo-cho, Matsuyama, Ehime 790-8577, Japan}

\author[0000-0001-9452-0813]{Takuma Izumi}
\affiliation{National Astronomical Observatory of Japan, 2-21-1 Osawa, Mitaka, Tokyo 181-8588, Japan}
\affiliation{Department of Astronomy, Graduate School of Science, The University of Tokyo, 7-3-1 Hongo, Bunkyo-ku, Tokyo 113-0033, Japan}
\affiliation{Graduate Institute for Advanced Studies, SOKENDAI, 2-21-1 Osawa, Mitaka, Tokyo 181-8588, Japan}
\affiliation{Kagoshima University, Kagoshima 890-0065, Japan}

\author[0000-0002-0106-7755]{Michael A. Strauss}
\affiliation{Department of Astrophysical Sciences, Princeton University, 4 Ivy Lane, Princeton, NJ 08544, USA}

\author[0000-0002-8434-880X]{Lilan Yang}
\affiliation{Laboratory for Multiwavelength Astrophysics, School of Physics and Astronomy, Rochester Institute of Technology, 84 Lomb Memorial Drive, Rochester, NY 14623, USA}

\author[0000-0003-3804-2137]{Knud Jahnke}
\affiliation{Max-Planck-Institut f\"{u}r Astronomie, K\"{o}nigstuhl 17, D-69117 Heidelberg, Germany}

\author[0000-0002-2099-0254]{Camryn L. Phillips}
\affiliation{Department of Astrophysical Sciences, Princeton University, 4 Ivy Lane, Princeton, NJ 08544, USA}

\author[0000-0002-8460-0390]{Tommaso Treu}
\affiliation{Department of Physics and Astronomy, University of California Los Angeles, CA, 90095, USA}

\author[0000-0001-6102-9526]{Irham T. Andika}
\affiliation{Technical University of Munich, TUM School of Natural Sciences, Department of Physics, James-Franck-Str. 1, D-85748 Garching, Germany}
\affiliation{Max-Planck-Institut f\"{u}r Astrophysik, Karl-Schwarzschild-Str. 1, D-85748 Garching, Germany}

\author[0000-0003-4569-1098]{Kentaro Aoki}
\affiliation{Subaru Telescope, National Astronomical Observatory of Japan, 650 North A'ohoku Place, Hillo HI 96720 USA}

\author[0009-0007-0864-7094]{Junya Arita}
\affiliation{Department of Astronomy, Graduate School of Science, The University of Tokyo, 7-3-1 Hongo, Bunkyo-ku, Tokyo 113-0033, Japan}

\author[0000-0002-9850-6290]{Shunsuke Baba}
\affiliation{Institute of Space and Astronautical Science (ISAS), Japan Aerospace Exploration Agency (JAXA), 3-1-1 Yoshinodai, Chuo-ku, Sagamihara, Kanagawa 252-5210, Japan}

\author[0000-0001-8582-7012]{Sarah E. I. Bosman}
\affiliation{Institute for Theoretical Physics, Heidelberg University, Philosophenweg 12, D-69120, Heidelberg, Germany}
\affiliation{Max-Planck-Institut f\"{u}r Astronomie, K\"{o}nigstuhl 17, 69117 Heidelberg, Germany}

\author[0000-0003-2895-6218]{Anna-Christina Eilers}
\affiliation{MIT Kavli Institute for Astrophysics and Space Research, Massachusetts Institute of Technology, Cambridge, MA 02139, USA}

\author[0000-0001-7201-5066]{Seiji Fujimoto}
\affiliation{Department of Astronomy, The University of Texas at Austin, Austin, TX 78712, USA}
\affiliation{David A. Dunlap Department of Astronomy and Astrophysics, University of Toronto, 50 St. George Street, Toronto, Ontario, M5S 3H4, Canada}

\author[0000-0003-3633-5403]{Zoltan Haiman}
\affiliation{Institute of Science and Technology Austria (ISTA), Am Campus 1, Klosterneuburg 3400, Austria}
\affiliation{Department of Astronomy, Columbia University, New York, NY 10027, USA}
\affiliation{Department of Physics, Columbia University, New York, NY 10027, USA}

\author[0000-0001-6186-8792]{Masatoshi Imanishi}
\affiliation{National Astronomical Observatory of Japan, Osawa, Mitaka, Tokyo 181-8588, Japan}

\author[0000-0001-9840-4959]{Kohei Inayoshi}
\affiliation{Kavli Institute for Astronomy and Astrophysics, Peking University, Beijing 100871, China}

\author[0000-0002-4923-3281]{Kazushi Iwasawa}
\affiliation{Institut de Ci\`encies del Cosmos (ICCUB), Universitat de Barcelona (IEEC-UB), Mart\'i i Franqu\`es, 1, 08028 Barcelona, Spain}
\affiliation{ICREA, Pg Llu\'is Companys 23, 08010 Barcelona, Spain}

\author[0000-0001-9187-3605]{Jeyhan Kartaltepe}
\affiliation{Laboratory for Multiwavelength Astrophysics, School of Physics and Astronomy, Rochester Institute of Technology, 84 Lomb Memorial Drive, Rochester, NY 14623, USA}

\author[0000-0003-3954-4219]{Nobunari Kashikawa}
\affiliation{Department of Astronomy, Graduate School of Science, The University of Tokyo, 7-3-1 Hongo, Bunkyo-ku, Tokyo 113-0033, Japan}
\affiliation{Research Center for the Early Universe, Graduate School of Science, The University of Tokyo, 7-3-1 Hongo, Bunkyo-ku, Tokyo 113-0033, Japan}

\author[0000-0002-3866-9645]{Toshihiro Kawaguchi}
\affiliation{Graduate School of Science and Engineering, University of Toyama, Gofuku 3190, Toyama 930-8555, Japan}

\author[0000-0002-1605-915X]{Junyao Li}
\affiliation{Department of Astronomy, University of Illinois at Urbana-Champaign, Urbana, IL 61801, USA}

\author[0000-0003-1700-5740]{Chien-Hsiu Lee}
\affiliation{Hobby-Eberly Telescope, McDonald Observatory, UT Austin, 32 Mount Fowlkes, Fort Davis, TX 79734, USA}

\author[0000-0001-6106-7821]{Alessandro Lupi}
\affiliation{Como Lake Center for Astrophysics,  DiSAT, Università degli Studi dell’Insubria, via Valleggio 11, I-22100, Como, Italy}

\author[0000-0002-4544-8242]{Jan-Torge Schindler}
\affiliation{Hamburger Sternwarte, University of Hamburg, Gojenbergsweg 112, D-21029 Hamburg, Germany}

\author[0000-0001-7825-0075]{Malte Schramm}
\affiliation{Universit\"{a}t Potsdam, Karl-Liebknecht-Str. 24/25, D-14476 Potsdam, Germany}

\author[0000-0002-2597-2231]{Kazuhiro Shimasaku}
\affiliation{Department of Astronomy, Graduate School of Science, The University of Tokyo, 7-3-1 Hongo, Bunkyo-ku, Tokyo 113-0033, Japan}
\affiliation{Research Center for the Early Universe, Graduate School of Science, The University of Tokyo, 7-3-1 Hongo, Bunkyo-ku, Tokyo 113-0033, Japan}

\author[0000-0002-7087-0701]{Marko Shuntov}
\affiliation{Cosmic Dawn Center (DAWN), Denmark}
\affiliation{Niels Bohr Institute, University of Copenhagen, Jagtvej 128, DK-2200, Copenhagen, Denmark}

\author[0009-0003-4742-7060]{Takumi S. Tanaka}
\affiliation{Kavli Institute for the Physics and Mathematics of the Universe (WPI), The University of Tokyo
Institutes for Advanced Study, The University of Tokyo, Kashiwa, Chiba 277-8583, Japan}
\affiliation{Department of Astronomy, Graduate School of Science, The University of Tokyo, 7-3-1 Hongo, Bunkyo-ku, Tokyo 113-0033, Japan}
\affiliation{Center for Data-Driven Discovery, Kavli IPMU (WPI), UTIAS, The University of Tokyo, Kashiwa, Chiba 277-8583, Japan}

\author[0000-0002-3531-7863]{Yoshiki Toba}
\affiliation{Department of Physical Sciences, Ritsumeikan University, Kusatsu, Shiga 525-8577, Japan}
\affiliation{National Astronomical Observatory of Japan, Osawa, Mitaka, Tokyo 181-8588, Japan}
\affiliation{Academia Sinica Institute of Astronomy and Astrophysics, 11F of Astronomy-Mathematics Building, AS/NTU, No.1, Section 4, Roosevelt Road, Taipei 10617, Taiwan}
\affiliation{Research Center for Space and Cosmic Evolution, Ehime University, 2-5 Bunkyo-cho, Matsuyama, Ehime 790-8577, Japan}

\author[0000-0002-3683-7297]{Benny Trakhtenbrot}
\affiliation{School of Physics and Astronomy, Tel Aviv University, Tel Aviv 69978, Israel}
\affiliation{Max-Planck-Institut f{\"u}r extraterrestrische Physik, Gie\ss{}enbachstra\ss{}e 1, 85748 Garching, Germany}
\affiliation{Excellence Cluster ORIGINS, Boltzmannsstra\ss{}e 2, 85748, Garching, Germany}

\author[0000-0003-1937-0573]{Hideki Umehata}
\affiliation{Institute for Advanced Research, Nagoya University, Furocho, Chikusa, Nagoya 464-8602, Japan}
\affiliation{Department of Physics, Graduate School of Science, Nagoya University, Furocho, Chikusa, Nagoya 464-8602, Japan}

\author[0000-0001-9191-9837]{Marianne Vestergaard}
\affiliation{DARK, The Niels Bohr Institute, University of Copenhagen, Jagtvej 155, 2200 Copenhagen N, Denmark}
\affiliation{Department of Astronomy and Steward Observatory, University of Arizona, 933 N Cherry Avenue, Tucson, AZ 85721, USA}

\author[0000-0002-7633-431X]{Feige Wang}
\affiliation{Department of Astronomy, University of Michigan, 1085 S. University Ave., Ann Arbor, MI 48109, USA}

\author[0000-0001-5287-4242]{Jinyi Yang}
\affiliation{Department of Astronomy, University of Michigan, 1085 S. University Ave., Ann Arbor, MI 48109, USA}

\begin{abstract}
The advent of JWST has opened new horizons in the study of quasar host galaxies during the reionization epoch ($z > 6$). Building upon our previous initial measurements of stellar light from two quasar host galaxies at these redshifts, we now report the detection of the stellar light from the full Cycle 1 sample of 12 distant moderate-luminosity quasar ($M_{1450}>-24$ mag) host galaxies at $z>6$ from the Hyper Suprime-Cam Subaru Strategic Program (HSC-SSP). Using JWST/NIRCam observations at 1.5 and 3.6 $\mu$m combined with 2D image decomposition analysis, we successfully detect the host galaxies in 11 of the 12 targets, underscoring the high detection rates achievable with moderate-luminosity quasars. Based on two-band photometry and SED fitting, we find that our host galaxies are massive, with log~M$_*$/M$_{\odot} = 9.5 \text{--} 11.0$. The effective radii range from 0.6 to 3.2 kpc, comparable to the sizes of inactive galaxies with similar masses at $z\sim6$ as measured with imaging from COSMOS-Web.
Intriguingly, the two quasar hosts with post-starburst features, which reside at the high-mass end of our sample and exhibit relatively compact morphologies, have similar size and stellar mass surface densities to quiescent galaxies at $z\sim\text{4--5}$.
These findings suggest that the so-called galaxy compaction scenario is already in place at the reionization epoch, in which gas inflows during starburst phases drive centrally concentrated star formation followed by rapid quenching, bridging the structural transition of massive galaxies from relatively extended star-forming disks to compact quiescent systems.
\end{abstract}



\section{Introduction} \label{sec:intro}

The formation of supermassive black holes (SMBHs) in the early universe, particularly at redshifts greater than 6, presents a major puzzle in modern astrophysics. These mysterious objects grew to an enormous scale of mass (e.g., $10^8-10^{10}\,\rm{M}_\odot$) in a relatively short period of cosmic history, thus challenging our understanding of black hole growth and galaxy evolution~\citep{Fan2006AJ....132..117F, Wu2015Natur.518..512W, Wang2021ApJ...907L...1W}.


The growth of SMBHs is linked to that of host galaxies through gas accretion. This process is regulated by multiple factors, such as the amount of cold gas available within the hosts, star formation activity, and the hosts' merger history.
AGN feedback may quench star formation or trigger compact star formation\footnote{Compact star formation refers to the concentration of star-forming activity in smaller, high-density regions, often associated with the buildup of stellar mass in the galaxy's core. See Section~\ref{diss:compaction} for a overview of this process.}~\citep{Zolotov2015MNRAS.450.2327Z, Inayoshi2020ARA&A..58...27I}. 
Constraining the properties of quasar host galaxies is particularly important at high redshift, as one can distinguish different theoretical scenarios for early black hole growth~\citep{Lupi2019MNRAS.488.4004L, Habouzit2021MNRAS.503.1940H}.
Among the most fundamental questions is how galaxies and their SMBHs established the tight correlation between stellar mass (or stellar velocity dispersion) and SMBH mass observed in the local universe~\citep{Kormendy2013ARA&A..51..511K}. 

Quasar hosts are also key to understanding the morphological transition of galaxies. 
The so-called ``compaction models" predict that extreme gas inflows, often triggered by mergers or disk instabilities, induce compact star formation and subsequent AGN activities, ultimately transforming disk galaxies into elliptical galaxies~\citep{Zolotov2015MNRAS.450.2327Z}.
Submillimeter observations, primarily from the Atacama Large Millimeter/submillimeter Array (ALMA), have revealed diverse gas dynamics among high-$z$ quasar hosts.
They show that about one-third of $z>6$ quasar hosts exhibit rotating gas disks, while another third show merger signatures~\citep{Decarli2019ApJ...880..157D, Neeleman2021ApJ...911..141N, Wang2024ApJ...968....9W}. 
Dynamical mass measurements further suggest that luminous quasar hosts have overmassive black holes with respect to the local stellar mass--SMBH mass relation~\citep{Izumi2019PASJ...71..111I, Izumi2021ApJ...914...36I}, whereas moderate-luminosity quasars appear less extreme~\citep{silverman2025shellqs}. However, these observations primarily trace cold gas and dust, which may not fully represent the stellar mass distribution or quiescent populations. 
Detecting host stellar light is thus essential to link these gas reservoirs to star formation and test consistency between tracers.

Observational studies of quasar host galaxies have been challenging, as the host stellar emission is often outshone by the quasar radiation, particularly at high redshift. At earlier epochs, the intrinsic sizes of galaxies appear to be smaller than at $z\sim0$~\citep{van_der_Wel2014, Shibuya2015ApJS..219...15S, Yang2022ApJ...938L..17Y}, which, combined with spatial resolution limits, making it difficult to distinguish them from the bright quasar cores~\citep{Zakamska2019}. Due to cosmic expansion, the stellar light from these galaxies undergoes significant redshifting, shifting its emission to longer wavelengths. Additionally, surface brightness dimming, which scales as $(1+z)^4$, significantly reduces their apparent brightness. These combined effects pose substantial challenges in detecting light above the rest-frame 4000 \AA, which typically encompasses the bulk of emission from stars.
For low-redshift quasars at $z<1$, ground-based surveys, such as Hyper Suprime-Cam Subaru Strategic Program \citep[HSC-SSP,][]{Aihara2018}, have been used to measure the structural and photometric properties of the host galaxies of $\sim$5000 Sloan Digital Sky Survey quasars~\citep{Li2021ApJ...918...22L}.
For quasars at $z\sim1\text{--}2$, HST studies have successfully detected hosts but faced limitations in resolving sub-galactic structures or separating AGN-dominated light in compact systems~\citep{Zakamska2019, Ding2020}. At $z>6$, even HST observations have struggled to detect host galaxy emission in most quasars~\citep{Mechtley2012ApJ...756L..38M, Marshall2020ApJ...900...21M}, a challenge attributed to the overwhelming contrast in flux between the quasars and their hosts, unless the system is undergoing a merger~\citep{Decarli2019ApJ...880..157D}, with the host galaxy stretched well beyond the tails of the PSF.

The deployment of the James Webb Space Telescope ~\citep[JWST;][]{Ribgy2023PASP..135d8001R} has marked a turning point in this area of study. 
Its large 6.5-meter mirror enables superior angular resolution and a high-quality point spread function (PSF), resulting in sharp imaging. Its position at the second Lagrange point further ensures thermal stability, maintaining a more stable PSF. These capabilities allow for the reliable decomposition of quasar and host galaxy emission.
Additionally, the NIRCam's infrared wavelength coverage at $1\text{--}5~\mu$m provides access to rest-frame optical emission from galaxies at high redshift, well beyond the limitation of HST studies. 

Early JWST studies have demonstrated its capability to detect quasar host galaxies at $z>6$.
In our previous work \citep{Ding2023, Onoue2024}, we reported robust detections of stellar light from three $z>6$ quasar hosts using NIRCam broadband images (F356W/F150W).
These hosts are found to be massive (log~M$_*$/M$_{\odot} = 10.5 \text{--} 11.0$) and compact (effective radius, \reff, of $0.5 \text{--} 2.0$ kpc). 
Stellar light is even resolved for the most luminous quasars ($M_{1450}\lesssim-26$):
The EIGER collaboration \citep{Yue2024} reported the detection of compact host galaxies (\reff$=1.6 \text{--}2.2$ kpc) for three luminous quasars using multi-band NIRCam imaging (F115W/F200W/F356W).
\citet{Stone2024} derived even smaller effective radii (1$\text{--}$1.5 kpc) from PSF-subtracted radial profiles of five $5\lesssim z\lesssim 7$ quasars. 
Diverse morphologies are observed by JWST at intermediate redshifts ($2<z<5$), where decomposition studies detect spiral arms, bars, clumps, and merger signatures in quasar hosts \citep{Ding2022,Kocevski2023ApJ...946L..14K, Tanaka2024, Zhuang2024ApJ...962...93Z}.

The access to the stellar light of quasar hosts enables direct comparisons of the stellar mass--SMBH mass ratios with those measured at lower redshifts~\citep{silverman2025shellqs}.
Recent JWST studies have identified a population of faint AGNs ($M_{\rm UV}\gtrsim-22$) at $z>4$, whose black hole masses lie above the local stellar mass--SMBH mass relation~\citep[e.g.,][]{Pacucci2023ApJ...957L...3P, Harikane2023ApJ...959...39H, Ubler2023A&A...677A.145U, Maiolino2024A&A...691A.145M}.
At the extreme end, the luminous quasar hosts also indicate that their SMBHs are overmassive, consistent with the ALMA dynamical mass estimates~\citep{Yue2024, Stone2023ApJ...953..180S, Stone2024}.
These apparent offsets in black hole mass relative to stellar mass have been interpreted as evidence of exotic early growth channels of SMBHs, such as heavy seed BHs or super-Eddington accretion. 
On the other hand, \citet{Li2025ApJ...981...19L} argue that the coupled effect of selection biases (i.e., finite detection limits and requirements on detecting broad lines) and measurement uncertainties can largely account for the reported offset of black hole masses, leaving the understanding of the intrinsic mass distribution at high redshift elusive.
Intriguingly, observations of moderate-luminosity ($-24\lesssim M_{1450}\lesssim-22$) quasars at $z>6$ , which are also affected by such biases, reveal host galaxies consistent with the local relation when accounting for selection effects~\citep{Ding2023, silverman2025shellqs}. 
In support of this conclusion, broad-line AGN have been confirmed from JWST among massive galaxies at $3<z<5$ \citep{Carnall2023, Li2025arXiv250205048L}.
These SMBHs are consistent with the local relation, highlighting the need for population studies that span a wide range of AGNs luminosities and host galaxy properties in order to disentangle true evolutionary trends from observational biases.

A large sample of $z>6$ quasars with secure host stellar light detections is crucial in this early stage of  JWST observations.
This paper presents the results of host galaxy decomposition and host galaxy properties for twelve quasars at $6.0 < z < 6.4$ in the moderate luminosity regime, building on the success of our first few targets reported in \citet{Ding2023} and \citet{Onoue2024}.
Our analysis of the mass relations between SMBHs and their host galaxies 
are presented in~\citep{silverman2025shellqs}.

This paper is organized as follows. In Section~\ref{sec:method}, we describe the JWST data and our sample selection. Section~\ref{sec:decomp} introduces our quasar image decomposition technique and the SED fitting. In Section~\ref{sec:result}, we present the results of our measurements and the size--mass relation compared to non-active galaxies from the COSMOS-Web catalog. Discussion and concluding remarks are presented in Section~\ref{sec:diss} and \ref{sec:sum}. Magnitudes are given in the AB system. The~\cite{Chabrier2003} initial mass function (IMF) is employed to infer the stellar mass of the host galaxies and the control sample. We use a standard concordance cosmology (i.e., $\Lambda$CDM model) with $H_0= 70$ km s$^{-1}$ Mpc$^{-1}$, $\Omega{_m} = 0.30$, and $\Omega{_\Lambda} = 0.70$.

\begin{table*}
    \centering
    \caption{Details of Target information}\label{table:target_info}
    \begin{tabular}{ccccccc}
\hline\hline
(1) & (2) & (3) & (4) & (5) & (6) & (7) \\
Target ID & RA & Dec & $z$ & Observing & $M_{1450}$ & $\#$ of PSF stars \\
& (hh mm ss.ss) & (dd mm ss.s) & & date & &(F150W, F356W) \\
\hline
HSC J2255+0251 & 22 55 38.04 & +02 51 26.6 & 6.34 & 2022-10-26 & $-$23.87 & 9, 5 \\
HSC J2236+0032 & 22 36 44.58 & +00 32 56.9 & 6.4 & 2022-11-05 & $-$23.75 & 16, 13 \\
HSC J0844$-$0132 & 08 44 08.61  & $-$01 32 16.5 & 6.18 & 2022-11-28 & $-$23.97  & 11, 14 \\
HSC J0911+0152 & 09 11 14.27 & +01 52 19.4 & 6.07 & 2022-12-25 & $-$22.09  & 14, 8 \\
HSC J0918+0139 & 09 18 33.17 & +01 39 23.4 & 6.19 & 2022-12-25 & $-$23.71  & 9, 8 \\
HSC J1425$-$0015 & 14 25 17.72 & $-$00 15 40.8 & 6.18 & 2023-02-27 & $-$23.44  & 13, 10 \\
HSC J1512+4422 & 15 12 48.71 & +44 22 17.5 & 6.19 & 2023-03-19 & $-$22.07  & 6, 4 \\
HSC J1525+4303 & 15 25 55.79 & +43 03 24.0 & 6.27 & 2023-03-19 & $-$23.61  & 7, 6 \\
HSC J1146$-$0005 & 11 46 58.90 & $-$00 05 37.6 & 6.3 & 2023-06-23 & $-$21.46  & 14, 5 \\
HSC J1146+0124 & 11 46 48.42 & +01 24 20.1 & 6.27 & 2023-06-23 & $-$23.71  & 7, 3 \\
HSC J0217$-$0208 & 02 17 21.59 & $-$02 08 52.6 & 6.2 & 2023-12-10 & $-$23.19 & 7, 3 \\
HSC J0844$-$0052 & 08 44 31.60 & $-$00 52 54.6 & 6.25 & 2023-11-16 & $-$23.74  & 18, 11 \\
\hline
    \end{tabular}
    \tablecomments{Column 1: Object IDs are listed in the full name format as originally defined in~\cite{Matsuoka2016}. For the remainder of this work, we will refer to each object by its short name (e.g., J2255+0251). Columns 2 and 3: J2000 R.A. and decl. coordinates. Column 4: spectroscopic redshift based on Ly$\alpha$ and Lyman break. Column 5: the JWST observing date. Column 6: The $M_{1450}$ value measured in~\cite{Matsuoka2016, Matsuoka2018, Matsuoka2018PASJ...70S..35M}. 
    Column 7: The number of PSF stars collected from the corresponding JWST NIRCam field of view in F150W and F356W.}
    \label{tab:my_label}
\end{table*}

\section{Experimental Design} \label{sec:method}
\subsection{Sample selection}

\begin{figure}
\centering
{\includegraphics[trim = 0mm 0mm 0mm 0mm, clip, height=0.45\textwidth]{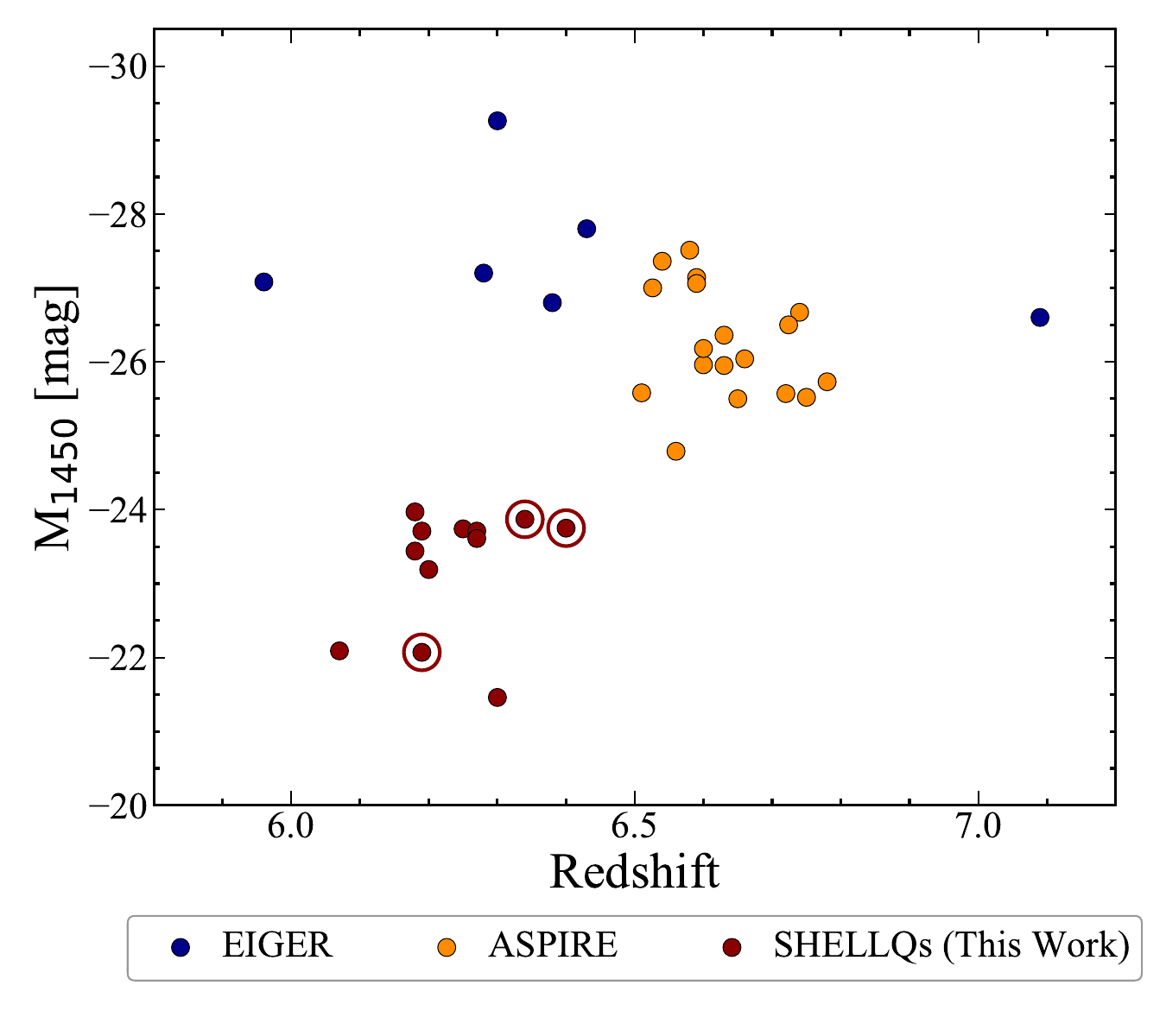}}
\caption{
\label{fig:qso_distribute}
The redshift-$M_{1450}$ distribution of the SHELLQs quasars in this study (red) and other major JWST observations for luminous quasars: EIGER (blue; \citealt{Yue2024}) and ASPIRE (orange; \citealt{Wang2023ApJ...951L...4W}).
The SHELLQs quasars, whose host detections were reported in our earlier papers \citep{Ding2023, Onoue2024} are indicated by larger open circles.
}
\end{figure}

The twelve quasars presented in this study were observed as part of a JWST General Observers (GO) program in Cycle~1 (GO-1967; PI: M. Onoue).
The basic properties of these quasars are summarized in Table~\ref{tab:my_label}.
Their redshifts and rest-frame 1450 \AA\ absolute magnitudes are presented in Figure~\ref{fig:qso_distribute}.
These quasars at $6.0<z<6.4$ were identified by the ``Subaru High-z Exploration of Low-Luminosity Quasars''~\cite[SHELLQs;][]{Matsuoka2016, Matsuoka2018, Matsuoka2018PASJ...70S..35M, Matsuoka2022} project through the HSC-SSP~\citep{Aihara2018}, a $1,100~\mathrm{deg}^2$-class optical survey conducted by the 8.2-m Subaru Telescope. 
The rest-frame 1450 \AA\ magnitudes of the targets are $-23.9 \leq M_{1450} \leq -21.5$ mag, approximately one order of magnitude fainter than those typically found in other major quasar surveys such as the Sloan Digital Sky Survey~\citep{Jiang2016ApJ...833..222J} and the Pan-STARRS1 survey~\citep{Banados2016ApJS..227...11B, Banados2023ApJS..265...29B}.
We selected these JWST targets mostly from a sample used in a study of the $z\sim6$ quasar luminosity function \citep{Matsuoka2018ApJ...869..150M} with two additional targets (J1146$-$0005 and J0911$+$0152) from the faintest SHELLQs quasars at the time of preparing the JWST proposal.
We also note that J0217$-$0208 shows a relatively narrow Ly$\alpha$ emission for a type-I quasar (FWHM$_\mathrm{Ly\alpha} <230$ km s$^{-1}$; \citealt{Matsuoka2018PASJ...70S..35M}). 
This object is among the sub-population of SHELLQs quasars, which we call ``narrow-line quasars", whose Ly$\alpha$ luminosities are greater than $L_\mathrm{Ly\alpha} = 10^{43}\ \mathrm{[erg\ s^{-1}]}$ and Ly$\alpha$ full widths at half maximum are narrower than 500 km s$^{-1}$ \citep{Matsuoka2022}.
While we regard J0217$-$0208 as a quasar in this study, its nature, based on rest-frame optical spectroscopy, is discussed in other works~\citep[e.g.,][]{Matsuoka2025ApJ...988...57M}.

There are advantages to observing  SHELLQs quasars for host galaxy studies~\citep{Fan2023ARA&A..61..373F}.
As compared to more luminous quasars (Figure~\ref{fig:qso_distribute}), such as those observed by the EIGER \citep{Yue2024} and ASPIRE \citep{Wang2023ApJ...951L...4W} projects, the ratio of stellar emission to total rest-frame optical emission is expected to be higher in our sample; this results in a lower contrast between the host galaxy and quasar emission, thereby enhancing the detectability of the host.
While even fainter AGN detected by JWST \citep[e.g.,][]{Onoue2023, Kocevski2024arXiv240403576K, Maiolino2024A&A...691A.145M, Akins2024arXiv240610341A, Juodvzbalis2025arXiv250403551J} may be even more effective for host detection, robust decomposition of AGN and host galaxy emission remains challenging due to their insufficient signal-to-noise ratios, leading to  
large uncertainties in stellar mass estimates. Therefore, the SHELLQs quasars appear to hit the sweet spot for host galaxy studies with JWST.

\subsection{JWST observation and data reduction}

The data were obtained using Module B of the NIRCam instrument, which provides a field of view of \(2.2 \times 2.2\) arcmin\(^2\). We obtained images with total exposures of $3,100$~s per target across the two filters (F356W and F150W), observed simultaneously. To correct for cosmic ray hits and bad pixels in the detector, and to facilitate subpixel resampling during the stacking phase, a \(4\times4\) primary and subpixel dithering pattern was employed. The INTRAMODULEBOX and STANDARD dithering patterns were utilized for primary and subpixel dithers, respectively. The BRIGHT1 readout mode was chosen for these observations. Note that this program also includes spectroscopic observations using NIRSpec Fixed Slit (S200A2), which enables the calibration of the SMBH mass~\citep[e.g.,][]{Ding2023, Onoue2024}. 
The analysis of spectroscopic results will be presented in forthcoming works (Onoue et al., in prep), while we focus on the imaging study in this paper.

We perform the data processing of the NIRCam images following the standard procedures outlined in the JWST pipeline v.1.7.2.\footnote{\url{https://jwst-crds.stsci.edu}} We downloaded the precalibrated `Stage 2' image frames from the MAST archive, which were processed with pipeline parameter reference files 
as registered in the JWST Calibration Reference Data System (CRDS). The initial steps involved subtracting global background light from individual frames using the Background2D function from 
{\sc photutils}~\citep{Photutils2016ascl.soft09011B}. 
Notably, the archived images exhibit horizontal and vertical stripe noise patterns, recognized as `1/f noise'. This noise was mitigated by first masking bright objects and then collapsing the 2D images along each axis of the detectors to estimate the noise amplitudes through sigma-clipped median values. We then subtracted the amplitudes from each row and column, with horizontal stripes measured for each of the four detector amplifiers separately.

We follow the standard approach and utilize the post-processed Stage 2 image frames for alignment and stacking using inverse-variance weighting in the Stage 3 standard pipeline. We preserved the original detector positions during this step, which ensures that the relative position angles of all fields of view remain consistent across the sample. This consistency is crucial for constructing the point spread function (PSF) library (see next section). Both F356W and F150W images were resampled to a pixel scale half that of the original detector pixel scale, utilizing the drizzling algorithm implemented in the resampling step of the pipeline. The resulting pixel scales for F356W and F150W images are $0\farcs0315$ and $0\farcs0153$, respectively. These pixel scales correspond to approximately $30\%$ of the full width at half maximum (FWHM) of the point spread function in each filter.

Having obtained the final science image, we perform a second round of background light removal to eliminate residual contributions from both the sky and the detector. For this purpose, we adopt the {\sc photutils} package~\citep{Photutils2016ascl.soft09011B}, which implements a two-dimensional background modeling based on the {\sc SExtractor} algorithm~\citep{SExtractor1996A&AS..117..393B}. This approach effectively accounts for gradients and variations across the background. Once the global background map is derived, we subtract it from our science frames to obtain cleaner images for subsequent analysis. We measure the surface brightness in the empty regions to verify that it is consistent with zero within the noise. This background removing process has been successfully used in our previous work~\citep[e.g.,][]{Ding2020, Ding2022, Ding2023}.

\subsection{Control sample from COSMOS-Web}

We establish a control sample of massive galaxies at $z\sim6$ to allow for a meaningful comparison with our quasar host galaxies in terms of size--mass relationships and morphological properties. To select a sufficient number of massive galaxies at high redshift, we utilize the catalog from the COSMOS-Web survey~\citep{Casey2023}, the largest JWST Cycle 1 treasury program. COSMOS-Web images a contiguous 0.54 deg$^2$ area with NIRCam and 0.19 deg$^2$ with MIRI in the COSMOS field~\citep{Scoville2007}. The survey employs four NIRCam filters (F115W, F150W, F277W, and F444W) reaching approximate 5-$\sigma$ depths of 27.4-28.2 AB mag, and one MIRI filter (F770W) with a 5-$\sigma$ depth of 26 AB mag.

The data reduction process for both NIRCam and MIRI observations are detailed in~\citep[][respectively]{Franco2025arXiv250603256F, Harish2025arXiv250603306H}, and we provide a brief overview here. The NIRCam raw imaging data were processed using the JWST Calibration Pipeline version 1.12.1, with additional custom modifications to address issues such as `1/f' noise and sky background subtraction, similar to other JWST studies~\cite[e.g.,][]{Bagley2023ApJ...946L..12B}. The Calibration References Data System file pmap.1170 is employed, which corresponds to NIRCam instrument mapping (imap) 0273. The final mosaics were created in Stage 3 of the pipeline with a pixel scale of 0\farcs03/pixel.

This control sample from COSMOS-Web provides us with a robust set of 422 massive star-forming galaxies at $z\sim6$ with log~M$_*$/M$_{\odot}$ from 9.0 to 10.5. We also include a sample of seven quiescent galaxies at $4<z<5$.
These stellar masses and photometric redshifts for this control sample are derived from the COSMOS catalog \citep{Shuntov2025arXiv250603243S} based on SED fitting, providing a well-characterized set of high-redshift galaxies.

The same \cite{Chabrier2003} IMF is adopted to infer the stellar mass for the COSMOS catalog. The size measurements for these control galaxies have been conducted and reported by \cite{Yang2025arXiv250407185Y} using the same methodology as our analysis of quasar hosts. In particular, \cite{Yang2025arXiv250407185Y} utilize the same image analysis tool, \galight~\citep[][see Section~\ref{subsec:decomp}]{Ding2020}, for galaxy fitting, ensuring a consistent approach in determining galaxy sizes across both the control and quasar host samples. This consistency is essential for enabling a fair and direct comparison of the size--mass relation between our quasar hosts and the general massive galaxy population at similar redshifts.

For the morphology comparison (e.g., section~\ref{subsec:SizeMass}), the COSMOS-Web control sample uses the F444W filter, while our quasar hosts are observed in the F356W filter. While these filters differ in central wavelength, both are long-wavelength bands (above rest-frame 4000~\AA) and are expected to trace similar stellar emission, ensuring a meaningful comparison of galaxy sizes and structures.


\section{Quasar-Host decomposition}\label{sec:decomp}
In this section, we outline the methodology used to decompose the quasar host galaxy light from the central nuclear light.

\subsection{PSF library}
\label{psf_lib}
The quality of the PSF model is essential for accurately characterizing the shape of the central point source (i.e., quasar) and separating it from its host, especially when the quasar dominates the overall emission. PSF variations occur temporally and spatially across the field of view of the detector due to aberration and breathing effects. 

In this study, we employ a methodology consistent with our previous works \citep{Ding2020, Ding2022, Ding2023, Tanaka2024} by constructing a PSF library. Previous studies ~\cite[e.g.][]{Zhuang2024} have examined the effectiveness of NIRCam's empirical PSFs generated by PSFEx~\citep{Bertin2011ASPC..442..435B}, which supports the modeling of spatial variations of the PSF model across the NIRCam's FoV as a function of pixel coordinates and provides a qualified PSF model to perform the AGN decomposition. As demonstrated by~\cite{Tanaka2024}, the PSF library approach produces consistent host inference results compared to these empirical PSFs by PSFEx. The details of the PSF library construction are described below.

For each target observed in the F150W and F356W bands, we utilize the PSF-star searching function within \galight\ to collect isolated, unsaturated PSF stars with adequate signal-to-noise ratios. We further improve the use of the selected PSF stars to produce a clean PSF model by erasing any nearby objects located in the PSF cutout --- replacing their pixels with values from empty (background) regions --- so they do not interfere with our analysis. We present the number of PSF stars selected for each target in Table~\ref{table:target_info}. Subsequently, each PSF in the library is applied during the fitting process, and the final result is obtained by combining the contributions from all PSFs, as detailed in Section \ref{weighting_stra}.

\begin{figure*}
    \centering
    \includegraphics[width=1\linewidth]{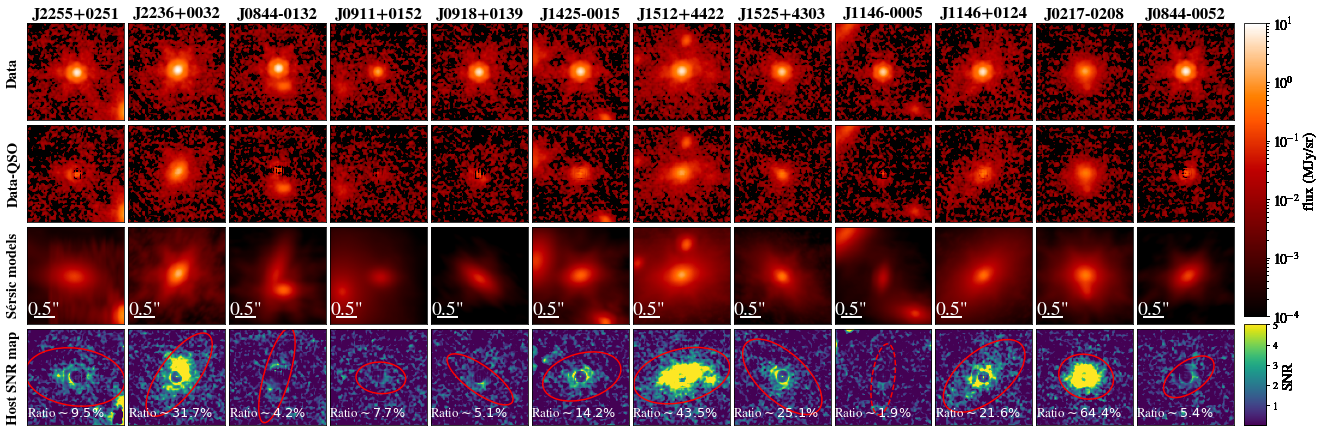}
    \caption{Quasar-host decomposition of NIRCam/F356W images for all twelve targets. The rows are as follows from top to bottom: (1) original science image (data), (2) data minus quasar model, i.e., the host galaxy and neighbors, (3) best-fit host \sersic\ model(s), and (4) Signal-to-noise ratio (SNR) maps of the host significance; the noise is a combination of that from the data and the scatter of the corresponding top-five PSFs. In Row (4), we show the elliptical apertures within which we calculate the host SNR, as reported in Table~\ref{tab:F356W_result}. When drawing these apertures, we use solid lines to highlight those with confirmed host detections. The inferred host-to-total flux ratios from the best fit are also indicated in the bottom panels. Our simulation results indicate that even in challenging cases, such as J0844$-$0132, we are able to accurately recover the host galaxy properties, as demonstrated in Figure~\ref{fig:sim_onecase}.}
    \label{fig:F356Wfit}
\end{figure*}

\begin{figure*}
    \centering
    \includegraphics[width=1\linewidth]{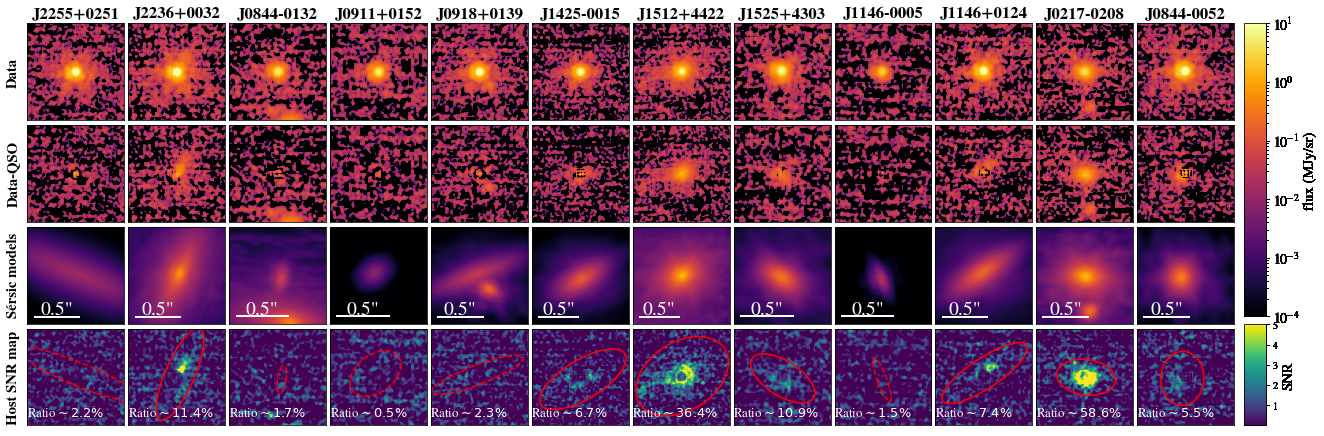}
    \caption{Same as Figure~\ref{fig:F356Wfit} for the F150W filter. Note that for non-significant detections (i.e., those indicated by dashed-line elliptical apertures), the inferred host properties have larger uncertainties. In these cases, the orientation of the inferred host shape in F150W can even be inconsistent with that in F356W.
    }
    \label{fig:F150Wfit}
\end{figure*}

\subsection{Image decomposition and inference of host properties}
\label{subsec:decomp}
We conduct two-dimensional modeling of images using our 2D profile fitting software \galight\footnote{The performance of \galight\ has been verified through simulations and comparisons with fitting by {\sc galfit} by \cite{Ding2023}.}. This is an open-source {\sc Python} package designed for astronomical data processing, which utilizes the image modeling capabilities in {\sc lenstronomy}~\citep{Birrer2018,Birrer2021}, a multi-purpose software package to model strong gravitational lenses. {\sc lenstronomy}'s flexibility enables \galight\ to turn off the lensing channel and focus on the AGN and host decomposition enhancing the user interface for automated fitting ability. Our modeling process involves the preparation of essential ingredients using \galight, including:
\begin{enumerate}
\item Science data image cutouts, covering sufficient quasar light emission. We set the cutout size as 81$\times$81 pixels (i.e., ~$2\farcs4\times2\farcs4$) for F356W and 71$\times$71 pixels (i.e., ~$1\farcs1\times1\farcs1$) for F150W, respectively. 
\item Noise level map, representing pixel value uncertainty.
\item PSF model, describing the point source shape and for image convolution. 
\end{enumerate}
The noise level encompasses both Poisson and random background noise components. The Poisson noise is estimated by calculating the effective exposure time based on weight maps (i.e., WHT map), accounting for gain values. The background rms noise level is determined using pixels from a blank region near the target. During the fitting, we do not explicitly include an additional noise term for the PSF model; instead, we account for PSF-related uncertainties by measuring the scatter in host galaxy properties obtained using different PSF models from our library, as described in Section~\ref{psf_lib}.

We assume that the central quasar is described as a scaled PSF at an arbitrary position and use a 2D \sersic\ profile~\citep{Sersic} to model the host galaxy. If any object happens to be close to our target, we add another \sersic\ profile to model its light and remove any potential contamination from its extended profile. To avoid any unphysical results, we limit the \sersic\ parameters as follows: effective radius \reff\ $\in [0\farcs06, 2\farcs00]$ (corresponding to a range of $\sim$0.33 to $\sim$11.16 kpc at $z\sim6.5$), \sersic\ index $n\in[1, 4]$.\footnote{While the light profile shape undergoes a dramatic transformation as the \sersic\ index value changes from 1 to 4, there is no such evident change in the light profile shape from 4 to higher values~\citep{Graham2013pss6.book...91G}.} The central position of the 2D \sersic\ model is allowed to be different from the central point source. Note that in \galight, the profiles of the central PSF and the galaxies are fitted simultaneously. Once the modeling is completed, we subtract the scaled point source component (i.e., quasar) from the image to isolate and assess the intrinsic emission from the host galaxy. This approach ensures a robust decomposition and avoids biases that may arise from sequential fitting. The final host flux is measured as the total light of the fitted \sersic\ model integrated within the image cutout region, which is typically consistent with the total flux calculated from the full \sersic\ profile. The fits to the F150W and F356W images are performed independently, with no shared or simultaneously fitted parameters between the two bands.

Note that the sizes of our host galaxies are defined using the \sersic\ effective radius (\reff) along the semi-major axis, which is consistent with the approaches adopted in the literature ~\citep[e.g.,][]{van_der_Wel2014} that are usually measured by {\sc galfit}~\citep{Peng2010}. For the final host photometry, we adopt the total flux of the \sersic\ model integrated within the image cutout region, ensuring that the host flux measurement is consistent with the area used in the decomposition analysis.

\subsection{Combining strategy}
\label{weighting_stra}
During the quasar image modelling task, there are different choices of fitting configurations that can be introduced, including the choice in PSF models, initial parameter settings, and the parameter minimization algorithm. These different choices can lead to variations in the inferred properties of quasar host galaxies, including flux and effective radius. To mitigate potential biases introduced by these variations, we employ a weighting approach that combines results from all configurations, ensuring robust and reliable measurements. This strategy allows us to quantify uncertainties arising from different fitting configurations, ensuring robust estimates of host galaxy properties while mitigating systematic biases.
We outline all the different fitting configurations or factors we considered that can cause the fitting results to vary, as follows:

\begin{itemize}
\item PSF supersampling factor: The parameter value of {\it point\_source\_supersampling} in \galight\ controls the interpolation factor for a sub-pixel shift of the PSF to align with the quasar's position. This interpolation within a sub-pixel can result in modest variations in the allocation of flux between the quasar and its host galaxy. To quantify any bias that related to this factor, we adopt PSF interpolation factor values of both 1 and 2.

\item \sersic\ index $n$: The $n$ value controls the central intensity concentration of the \sersic\ model, which inherently is degenerate with the flux value of the point source -- higher \sersic\ index values result in a more concentrated host profile, leading to increased inferred host flux. However, due to the compact nature of our host galaxies, accurately fitting this index value is challenging and can lead to potential bias. To account for this, we carried out the fits by fixing the $n$ to four typical values: 1, 2, 3, and 4.

\item PSF model: The accuracy of the PSF model is crucial in the decomposition task. In this study, we adopt each PSF from the entire PSF library for a fixed fit configuration (i.e., after setting the {\it point\_source\_supersampling} and $n$ values). For one fixed fitting configuration, we selected the top five PSFs based on their corresponding $\chi^2$ values, to ensure that only the most suitable PSFs contribute to the final combined result. As discussed by~\cite{Zhuang2024}, the smallest reduced $\chi^2$ does not necessarily indicate the best-fitting result. Therefore, we adopt a weighting strategy that combines these top PSFs evenly to mitigate any potential bias arising from PSF inaccuracies.
\end{itemize}

For a fixed target in a given broad-band image, the results are combined using the values derived from the 40 configurations (i.e., $2 \times4 \times5$: PSF supersampling factors $\times$ \sersic\ indices $\times$ PSF models, respectively). Our final inferred host properties are determined by calculating the mean value and standard deviation of their distribution across the 40 configurations. We assume that the true values of the host properties are covered within the random scatter of the results obtained from all available fitting configurations, as mentioned above. As will be shown later in Section~\ref{sec:result} (see Table~\ref{tab:F356W_result} and~\ref{tab:F150W_result}), the error bars associated with our host value are usually small. This indicates that the outcomes from these different settings do not vary excessively, thereby enhancing our confidence in our final estimates.

\subsection{Criteria for significant detection}\label{sec:detect_crit}
In the process of quasar image decomposition, incorporating a \sersic\ profile model alongside a point source model provides an inferred flux value for the extended component. However, this flux value alone is insufficient to confidently assert a robust detection of the host galaxy. To establish a reliable basis for claiming such detections, we have defined specific criteria for host galaxy identification in this study. For a target in the F150W or F356W filter, we design three key criteria that must be met to consider the quasar host galaxy as a {\it significant} detection:

\begin{itemize}
\item Sufficient host-to-total flux ratio ($>3\%$): The inferred host galaxy flux must contribute at least 3\% to the total flux (i.e., integrated model fluxes of the quasar and host galaxy). As demonstrated in Section~\ref{sec:simulation}, we have conducted extensive simulation tests to validate the robustness of our host-to-total $>3\%$ criterion. This criterion ensures that the detected host is substantial enough to be distinguished from noise or observational artifacts. 

\item Adequate signal-to-noise ratio (SNR$\geq$2) of the host.
A sufficiently high SNR ensures that the detected host signal is reliably above the noise floor. For each of the broad-band images F356W and F150W, we set SNR$\geq$2 as the criterion for significant detection.
In our later analysis, we find that almost all of the targets can meet this criterion in F356W.

\item Improved $\Delta$BIC$>$50: The addition of a \sersic\ profile to the model must result in an improved fit compared to a single point source alone, according to the Bayesian Information Criterion (BIC) comparison. 
\end{itemize}

For the second criterion, we adopt the fitting result and calculate the `host' SNR by comparing the flux of the data minus the point source model (i.e., the host galaxy flux) to the central noise level. 
Selecting an appropriate aperture size is essential to obtain a representative host SNR for assessing the presence of the quasar host. If the aperture is too small, the measurement becomes dominated by central noise from PSF subtraction, leading to a low SNR. In contrast, an overly large aperture encompasses many empty pixels at the outskirts, which increases the total noise and reduces the SNR. Thus, we adopt an aperture, matched in shape and size to the inferred host, that yields the maximum SNR, balancing the inclusion of genuine host signal against the impact of noise.
Note that this noise level incorporates both the instrumental noise and the scatter of the top five PSFs used in the analysis. This SNR is defined as the ratio between the total flux value to the total noise level within a given region.
Note that this total noise is defined by the square root of the sum of the squared noise values for each individual pixel. The total SNR for the host is calculated as:
\begin{equation}
    \text{SNR} = \frac{\sum \rm{flux}_i}{\sqrt{\sum \rm{noise}_i^2}},
    \label{eq:snr}
\end{equation}
where `i' corresponds to the pixel$_{\rm i}$ within the defined apertures.

In the third criterion, the BIC value is computed as 
\begin{equation}
\label{eq:bic}
{\rm BIC}={\rm ln}(N_d)N_k-2{\rm ln}(\hat{L}),
\end{equation}
where $N_d$  and $N_k$ are the number of data points and free parameters within the model, respectively. $\hat{L}$ is the maximum likelihood value given the model.
$\Delta$BIC is the value of the difference in BIC between two fits: one fit considers the quasar as just a PSF component, and the other fit has an extra \sersic\ model included. For objects that pass this requirement demonstrates, the \sersic\ component is necessary to explain the residual flux after subtracting the point source, indicating the presence of an extended structure, with a penalty for the larger number of parameters in the model to avoid overfitting. 

We conducted realistic simulations to validate our detection criteria, including thresholds for the host flux ratio and SNR, as detailed in Section~\ref{sec:simulation}. Our simulations show that when the host galaxy is not added in the simulation, the corresponding host-to-total flux ratio, SNR, and $\Delta$BIC consistently remain below 1\%, 0.5, and 10, respectively, failing to meet these detection thresholds simultaneously. These criteria ensure that our claimed detections are robust and minimize the risk of false positives due to fitting artifacts or noise fluctuations.

\section{Results} \label{sec:result}

\subsection{Host inference in F356W and F150W}
We now present the results of our decomposition routine to detect the host galaxies for the 12 SHELLQs quasars in F356W and F150W (Figure~\ref{fig:F356Wfit} and~\ref{fig:F150Wfit}, respectively). 
As illustrated in these figures, we detect the host galaxies for the majority of our quasars at a significant level: 11/12 and 7/12 for the F356W and F150W filters, respectively. 
The inferred values of the key parameters based on the top-40 configurations (see Section~\ref{weighting_stra}) are listed in Tables~\ref{tab:F356W_result} (F356W) and~\ref{tab:F150W_result} (F150W). 
It is encouraging that the results for J2255+0215 and J2236+0032 are consistent with those reported by ~\cite{Ding2023}, although we have updated our fitting strategy with a larger PSF library and a new combination method (Section~\ref{weighting_stra}). 
We also present the \sersic\ index and $\Delta$BIC value based on the best-fit configuration in Table~\ref{tab:best_n_BIC}. 
For targets classified as non-detections under our criteria (Section~\ref{sec:detect_crit}), we derive upper limits by taking the maximum host flux measured across the 40 fitting configurations. This approach accounts for PSF-related uncertainties and provides conservative estimates of undetected host emission.

The reduced host detection rate in F150W (rest frame $\approx2000$ \AA) can be attributed to three key factors. 
First, the quasar continuum has a relatively flat shape in $F_\nu$ space and thus is bright in UV, while the stellar emission from an old stellar population is brighter at rest-frame optical wavelengths, which are captured by the F356W images (centered on $\approx5000$ \AA\ rest-frame at $z\sim6$).
Second, F356W benefits from JWST's enhanced sensitivity at longer wavelengths, achieving a lower noise floor compared to F150W for identical exposure times. 
Third, dust extinction impacts the F150W band more significantly compared to F356W.
These effects combine to produce lower host SNR in F150W. As a result, fewer quasar host galaxies meet our detection criteria in F150W.

\begin{table*}
    \centering
    \caption{Summary of target decomposition result in F356W}
    \begin{tabular}{ccccccccc}
\hline\hline
Target ID & host ratio & host mag  & quasar mag & host \reff\ & host \reff\ & host $q$ & offset & host SNR  \\
&&(AB)&(AB)&($''$)&(kpc)& ($b/a$) &(kpc)& \\
(1) & (2) & (3) & (4) & (5) & (6) & (7) & (8) & (9)\\
\hline
J2255+0251&9.6\%$\pm$1.3\% & 24.56$\pm$0.14 & 22.12$\pm$0.03 & 0.48$\pm$0.20 & 2.68$\pm$1.09&0.43$\pm$0.10&0.47$\pm$0.11&8.3$\pm$1.2 \\ 
J2236+0032&28.9\%$\pm$7.8\% & 23.01$\pm$0.28 & 22.00$\pm$0.13 & 0.11$\pm$0.02 & 0.61$\pm$0.14&0.35$\pm$0.04&0.21$\pm$0.09&28.4$\pm$10.7 \\ 
J0844$-$0132&3.8\%$\pm$0.9\% & 25.70$\pm$0.34 & 22.13$\pm$0.01 & 0.56$\pm$0.27 & 3.15$\pm$1.49&0.26$\pm$0.07&0.46$\pm$0.25&2.0$\pm$0.5 \\ 
J0911+0152&9.7\%$\pm$2.7\% & 26.56$\pm$0.30 & 24.10$\pm$0.02 & 0.16$\downarrow$ & 0.89$\downarrow$&0.56$\pm$0.14&0.24$\pm$0.14&8.4$\pm$2.2 \\ 
J0918+0139&7.0\%$\pm$1.6\% & 25.28$\pm$0.25 & 22.44$\pm$0.02 & 0.23$\pm$0.07 & 1.30$\pm$0.39&0.61$\pm$0.23&0.14$\pm$0.14&4.2$\pm$1.0 \\ 
J1425$-$0015&18.1\%$\pm$4.9\% & 24.12$\pm$0.29 & 22.44$\pm$0.07 & 0.11$\pm$0.03 & 0.61$\pm$0.16&0.57$\pm$0.11&0.17$\pm$0.09&15.4$\pm$4.3 \\ 
J1512+4422&36.2\%$\pm$7.7\% & 23.10$\pm$0.28 & 22.46$\pm$0.11 & 0.19$\pm$0.04 & 1.05$\pm$0.23&0.53$\pm$0.04&0.09$\pm$0.12&58.8$\pm$20.4 \\ 
J1525+4303&19.2\%$\pm$4.0\% & 24.61$\pm$0.24 & 23.02$\pm$0.05 & 0.16$\pm$0.04 & 0.89$\pm$0.20&0.51$\pm$0.08&0.14$\pm$0.06&15.0$\pm$2.9 \\ 
J1146$-$0005&3.9\%$\downarrow$ & 26.38$\uparrow$ & 22.90$\pm$0.02 & 0.13$\pm$0.05 & 0.70$\pm$0.30&0.51$\pm$0.20&0.51$\pm$0.23&0.9$\pm$0.5 \\ 
J1146+0124&17.1\%$\pm$2.5\% & 24.15$\pm$0.16 & 22.43$\pm$0.03 & 0.37$\pm$0.12 & 2.04$\pm$0.65&0.44$\pm$0.05&0.15$\pm$0.09&21.5$\pm$3.3 \\ 
J0217$-$0208&69.2\%$\pm$9.8\% & 23.69$\pm$0.18 & 24.62$\pm$0.35 & 0.12$\pm$0.02 & 0.68$\pm$0.11&0.80$\pm$0.04&0.09$\pm$0.06&107.0$\pm$21.6 \\ 
J0844$-$0052&4.1\%$\pm$1.6\% & 25.57$\pm$0.51 & 22.03$\pm$0.02 & 0.23$\pm$0.14 & 1.27$\pm$0.79&0.45$\pm$0.17&0.31$\pm$0.16&2.3$\pm$0.8 \\ 
\hline
    \end{tabular}
    \tablecomments{The values presented in this table are the averaged results of our host galaxy inference, obtained through a comprehensive approach utilizing 40 fitting configurations (see Section~\ref{weighting_stra}), including different PSFs, different point source sampling factors, and a range of \sersic\ index values. Both the SNR values and their uncertainties in the table are derived from the distribution of SNRs measured over the 40 fitting configurations. J0911+0152 has the faintest quasar, which reduces the impact of PSF mismatch on the host decomposition. As a result, the derived host galaxy SNR for this target is high given its appearance in Figure~\ref{fig:F356Wfit}.
    Column (8) indicates the position offset between the quasar and the center of the inferred host galaxy.
    For J0911+0152 in F356W, we set 0\farcs16 as the upper limit for the \reff\ during the fitting. Note that for J1146$-$0005 (the only non-significant detection case in F356W), we report the maximum host flux ratio and the corresponding magnitude derived from the 40 fitting configurations with downward/upward arrows. This corresponding flux is adopted as a 1-$\sigma$ upper limit to perform the SED fitting (Section~\ref{subsec:sed}); the other host properties of J1146$-$0005 have high uncertainties.
    \label{tab:F356W_result}}
\end{table*}

\begin{table*}
    \centering
    \caption{Summary of target decomposition result in F150W}
    \begin{tabular}{ccccccccc}
\hline\hline
Target ID & host ratio & host mag  & quasar mag & host \reff\ & host \reff\ & host $q$ & offset & host SNR  \\
&&(AB)&(AB)&($''$)&(kpc)& ($b/a$) &(kpc)& \\
(1) & (2) & (3) & (4) & (5) & (6) & (7) & (8) & (9)\\
\hline
J2255+0251&3.0\%$\downarrow$ & 26.66$\uparrow$ & 22.87$\pm$0.03 & \nodata & \nodata &\nodata&\nodata& 1.5$\pm$0.3 \\ 
J2236+0032&11.2\%$\pm$1.4\% & 25.00$\pm$0.15 & 22.74$\pm$0.01 & 0.13$\pm$0.04 & 0.72$\pm$0.20&0.29$\pm$0.06&0.30$\pm$0.04&7.2$\pm$1.1 \\ 
J0844$-$0132&7.8\%$\downarrow$ & 26.79$\uparrow$ & 24.06$\pm$0.02 & \nodata & \nodata &\nodata&\nodata& 2.3$\pm$0.8 \\ 
J0911+0152&3.5\%$\downarrow$ & 27.61$\uparrow$ & 24.00$\pm$0.01 & \nodata & \nodata &\nodata&\nodata&1.3$\pm$0.6 \\ 
J0918+0139&6.2\%$\downarrow$ & 26.03$\uparrow$ & 23.05$\pm$0.01 & \nodata & \nodata &\nodata&\nodata&1.9$\pm$0.5 \\ 
J1425$-$0015&8.5\%$\pm$2.3\% & 25.78$\pm$0.29 & 23.17$\pm$0.02 & 0.13$\pm$0.05 & 0.73$\pm$0.27&0.39$\pm$0.10&0.13$\pm$0.12&4.0$\pm$1.1 \\ 
J1512+4422&40.2\%$\pm$5.5\% & 24.07$\pm$0.17 & 23.63$\pm$0.09 & 0.08$\pm$0.01 & 0.44$\pm$0.04&0.69$\pm$0.03&0.07$\pm$0.04&31.4$\pm$3.9 \\ 
J1525+4303&14.9\%$\pm$2.6\% & 25.60$\pm$0.21 & 23.69$\pm$0.03 & 0.07$\pm$0.01 & 0.38$\pm$0.07&0.57$\pm$0.16&0.09$\pm$0.03&11.0$\pm$2.1 \\ 
J1146$-$0005&5.2\%$\downarrow$ & 28.04$\uparrow$ & 24.83$\pm$0.02 & \nodata & \nodata &\nodata&\nodata&0.8$\pm$0.6 \\ 
J1146+0124&8.1\%$\pm$1.4\% & 25.62$\pm$0.20 & 22.96$\pm$0.01 & 0.18$\pm$0.06 & 1.03$\pm$0.35&0.30$\pm$0.05&0.16$\pm$0.08&4.2$\pm$0.8 \\ 
J0217$-$0208&60.9\%$\pm$5.0\% & 24.29$\pm$0.13 & 24.78$\pm$0.10 & 0.06$\pm$0.00 & 0.34$\pm$0.00&0.62$\pm$0.03&0.07$\pm$0.04&57.4$\pm$7.2 \\ 
J0844$-$0052&7.8\%$\pm$1.7\% & 25.34$\pm$0.25 & 22.64$\pm$0.02 & 0.07$\pm$0.01 & 0.39$\pm$0.06&0.55$\pm$0.18&0.33$\pm$0.06&4.2$\pm$1.1 \\ 
\hline
    \end{tabular}
    \tablecomments{Same format as Table~\ref{tab:F356W_result} but for F150W. For the non-significant detections listed in this table, we use the corresponding flux value to provide upper limit constraints in the SED fitting process (Section~\ref{subsec:sed}); their morphological parameters are not presented in this table.
    }
    \label{tab:F150W_result}
\end{table*}

\begin{table}
    \centering
    \caption{Target decomposition result based on best-fit configurations}
    \begin{tabular}{ccccr}
\hline\hline
Target ID & \multicolumn{2}{c}{host \sersic\ index} & \multicolumn{2}{c}{$\Delta$BIC} \\
\hline
 & F356W & F150W & F356W & F150W \\
\hline
J2255+0251&1 & \underline{1} & 6145.82 & \underline{21.81} \\ 
J2236+0032&4 & 4 & 53280.78 & 2170.16 \\ 
J0844$-$0132&1 & \underline{1} & 1011.21 & \underline{$-$36.72} \\ 
J0911+0152&1 & \underline{1} & 170.86 & \underline{$-$48.23} \\ 
J0918+0139&1 & \underline{1} & 1357.33 & \underline{29.40} \\ 
J1425$-$0015&1 & 1 & 6932.52 & 852.89 \\ 
J1512+4422&4 & 2 & 38697.94 & 9614.43 \\ 
J1525+4303&4 & 1 & 3885.35 & 441.37 \\ 
J1146$-$0005&\underline{1} & \underline{1} & \underline{35.50} & \underline{$-$47.06} \\ 
J1146+0124&4 & 1 & 9311.33 & 708.64 \\ 
J0217$-$0208&2 & 2 & 18754.58 & 7153.36 \\ 
J0844$-$0052&1 & 1 & 3240.59 & 1423.38 \\ 
\hline
    \end{tabular}
    \tablecomments{Table presents the values based on the fitting results using the lowest $\chi^2$ values among all the fitting configurations. Note that the \sersic\ $n$ values listed are based on the fixed parameters during fitting. Since the host light is often heavily contaminated by the central quasar, these index values are poorly constrained in the presence of bright nuclear emission. Therefore, these values should be regarded as reference estimates only and interpreted with caution. The $\Delta$BIC values are used as one of the criteria to determine the host detection~(Section~\ref{sec:detect_crit}). The parameters of the non-significant detections are indicated with underline.}
    \label{tab:best_n_BIC}
\end{table}

\begin{figure}
\centering
{\includegraphics[trim = 0mm 0mm 0mm 0mm, clip, height=0.45\textwidth]{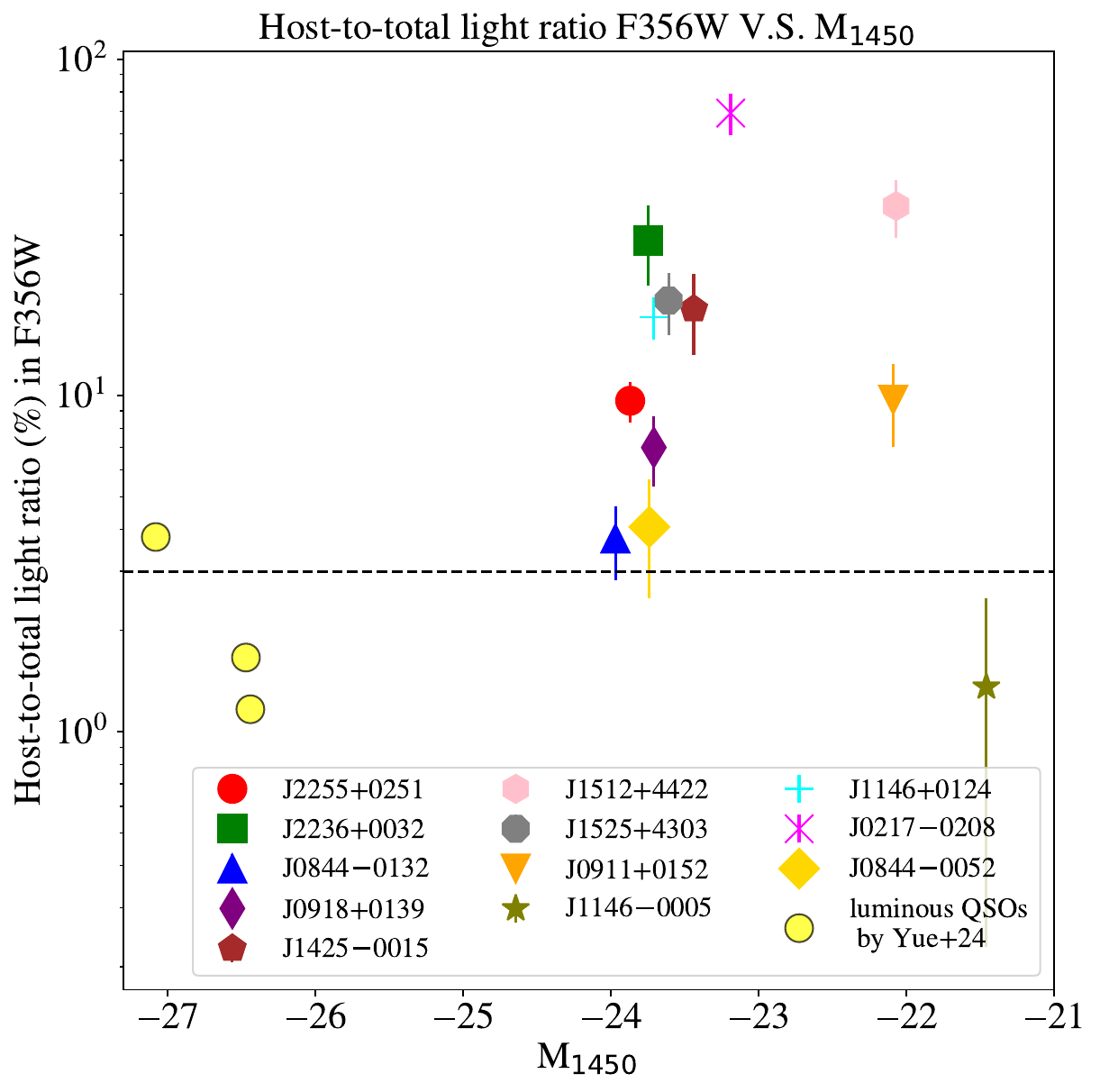}}
\caption{\label{fig:ratio}
Host-to-total flux ratio as a function of the quasar's absolute magnitude. The plot includes three luminous $z\sim6$ quasars reported in \cite{Yue2024}. Our results demonstrate that the intrinsically low luminosity of our quasars leads to a significantly higher host-to-total flux ratio than found in luminous quasars. Note that the photometric measurements of the host galaxies in \cite{Yue2024} are based on a different methodology that results in very small statistical uncertainties, leading to exceptionally small error bars in their sample. The 3\% threshold, which serves as one of our detection criteria for hosts, is indicated by the dashed line.
}
\end{figure}

Figure~\ref{fig:ratio} presents the distribution of the host-to-total flux ratio in F356W, as a function of AGN magnitude at rest-frame 1450 \AA.
The host-to-total flux ratio varies significantly across our sample, from $<4$ \% to 69 \%\ in the F356W images.
This wide variation in the stellar light contribution suggests that our UV-selected quasar sample spans a diversity of host galaxy populations.
Notably, the highest host fraction is found for J0217$-$0208 (the narrow-line quasar), suggesting that it has a distinct nature compared to the other quasars in our sample.
Also remarkable is that the two post-starburst galaxies reported in \citet{Onoue2024}, J1512+4422 and J2236+0032, exhibit the second and third highest stellar emission contrasts, respectively.
These two galaxies are likely the most mature, having already experienced their major starburst episodes.
Figure~\ref{fig:ratio} also shows the three successful host detections reported by \citet{Yue2024}, where the moderate-luminosity quasars in our sample exhibit significantly higher host-to-total flux ratios compared to the brighter counterparts ($<4$\%).
These higher stellar emission contrasts enable the high detection rate of host galaxies, a clear advantage of using moderate-luminosity quasars for studying quasar hosts.

We calculate the physical scale in kiloparsecs based on the effective radius of each host galaxy. In the F356W filter, the inferred host \reff\ (along the semi-major axis) ranges from 0.6 kpc to 3 kpc, while the sizes inferred in the F150W filter are typically smaller.  The host flux of J0911+0152 is weak in F356W, causing the inferred \reff\ to approach the upper limit of 2\farcs00, which is unphysical. To avoid any overestimation, we use the apparent size (i.e., the region exhibiting positive remaining flux) of the host image after subtracting the central point source (i.e.,  `Data$-$quasar' in Figure~\ref{fig:F356Wfit}) as an upper limit to perform the image fitting for this target. 
We also find that the inferred host galaxies are not circular: the ellipticities ($q = b/a$) are typically below 0.6.

In our work, the fits to the F150W and F356W images are performed independently. We note that the inferred host morphology in F150W, specifically the shape and orientation, sometimes differs from that in F356W. To assess the impact of these differences, we perform a test in which the $q$ and position angle (PA) in F150W are fixed to match the best-fit value from F356W. We found that this constraint resulted in negligible changes to the host photometry ($<0.05$ mag), indicating that our measurements of host flux are robust against uncertainties in morphological orientation between bands.

Our result also shows that, in general, our quasars exhibit positional offsets from their host centers of up to 0.5~kpc. These offsets may arise from host galaxy asymmetries, such as tidal distortions caused by galaxy interactions or clumpy accretion of cold gas. For example, ALMA observations of $z>6$ quasar hosts have revealed sub-kpc-scale offsets between the central black holes and the surrounding interstellar medium, often in interacting systems or those with close companions~\citep{Decarli2019ApJ...880..157D,Venemans2020}. Such offsets could also be caused by uneven dust attenuation in the host. With our targets at $z\sim6$, the F356W filter targets the rest-frame $g$-band, which can be affected by inhomogeneous dust attenuation, resulting in offsets between the quasar and stellar light. The offsets observed in some objects differ between the F150W and F356W filters. As an example, the offset directions in J0844$-$0052 appear to be nearly opposite in the two filters, as shown in Figure~\ref{fig:F356Wfit} and~\ref{fig:F150Wfit}. Such wavelength-dependent shifts are expected if the offset is due to obscuration.

Observations at longer wavelengths, such as JWST's NIRCam F444W and MIRI bands or (sub)millimeter wavelengths with ALMA, are required, to confirm the presence and spatial distribution of dust in these hosts. This would help determine whether dust attenuation is responsible for the suppressed rest-UV emission and lower detection rates in F150W, as more heavily obscured hosts may remain undetectable at shorter wavelengths but become visible in the rest-frame optical or far-infrared. However, we note that the observed host colors are not particularly red and can be well explained by normal stellar populations, suggesting that dust extinction may not be the primary driver of the lower detection rates in the UV.

We further investigate systematic uncertainties and perform image simulation tests to validate the accuracy of our host galaxy measurements; see Sections~\ref{sec:systematic} and~\ref{sec:simulation} for details.

\subsection{Multi-band host image fitting for J2236+0032\label{sec:J2236}}
\cite{Ding2023} confirmed that J2236+0032 ($z = 6.4$) has a massive extended host galaxy. As part of our ongoing investigation into high-redshift quasar host galaxies, we conducted follow-up observations of this quasar using JWST/NIRCam with six additional broad-band and medium filters, namely F115W, F200W, F250M, F300M, F444W, and F480M. This follow-up study (JWST-GO-3859 PI: Onoue) aimed to provide a comprehensive view of the quasar host galaxy's properties across a range of near-infrared wavelengths. Together with F150W and F356W, this multi-band approach allows us to probe various aspects of the host galaxy, from its stellar populations to its dust content, providing a more complete picture of the galaxy's structure and composition. The extended wavelength coverage offers unique insights into the dust-obscured star formation and enables more precise estimations of stellar properties.

The host decomposition of J2236+0032 was performed using the same approach as introduced in Section~\ref{subsec:decomp} and~\ref{weighting_stra}; the fit in each band is also considered as independent. The results for all the NIRCam filters are shown in Figure E1a of~\cite{Onoue2024} and in Table~\ref{table:J2236_result}. Our analysis indicates that the inferred host galaxy morphology is consistent across all eight bands, including the best-fit parameters \reff\ and $q$. This consistency not only suggests a robust and reliable morphological structure across different wavelengths but also indicates that the host is well described by a single-component model, with no evidence for a significant disk+bulge decomposition. As a test, we perform a joint fit of the eight-band data while applying a common \sersic\ profile with varying amplitudes for each band. 
The resulting host galaxy properties derived from this joint fit agree with those derived from independent fits, with any differences remaining within the $1\text{-}\sigma$ uncertainty range.

It is worth mentioning that our host decomposition results for J2236+0032 were presented in the study by~\cite{Onoue2024}, where we compare our NIRCam imaging-based decomposition with the spectral decomposition of the AGN and host galaxy flux within the NIRSpec slit. We found that our NIRCam image decomposition results are consistent with the spectral decomposition (see also Phillips et al. in prep.). This agreement between the two independent methods (i.e., imaging-based and spectral-based) provides additional validation for the state-of-the-art decomposition technique and enhances the reliability of the host galaxy properties in this work.


\begin{table*}
\centering
\caption{Summary of J2236+0032 decomposition results in eight NIRCam filters}
\begin{tabular}{ccccccc}
\hline\hline
Filter & host to total ratio & host mag  & quasar mag & host \reff\ & host \reff\ & host $q$  \\
&&&& ($''$)&(kpc)&  \\
\hline
F115W & 10.7\%$\pm$2.3\% & 25.45$\pm$0.23 & 23.12$\pm$0.03 & 0.14$\pm$0.04 & 0.80$\pm$0.23 & 0.42$\pm$0.05 \\
F150W & 11.2\%$\pm$1.4\% & 25.00$\pm$0.15 & 22.74$\pm$0.01 & 0.13$\pm$0.04 & 0.72$\pm$0.20 & 0.29$\pm$0.06 \\
F200W & 11.4\%$\pm$1.2\% & 24.49$\pm$0.12 & 22.26$\pm$0.01 & 0.18$\pm$0.04 & 0.98$\pm$0.22 & 0.45$\pm$0.02 \\
F250M & 13.2\%$\pm$4.9\% & 24.22$\pm$0.40 & 22.11$\pm$0.05 & 0.15$\pm$0.07 & 0.82$\pm$0.37 & 0.34$\pm$0.09 \\
F300M & 26.9\%$\pm$5.1\% & 23.15$\pm$0.21 & 22.05$\pm$0.08 & 0.12$\pm$0.02 & 0.65$\pm$0.11 & 0.37$\pm$0.01 \\
F356W & 28.9\%$\pm$7.8\% & 23.01$\pm$0.28 & 22.00$\pm$0.13 & 0.11$\pm$0.02 & 0.61$\pm$0.14 & 0.35$\pm$0.04 \\
F444W & 31.3\%$\pm$8.4\% & 22.79$\pm$0.29 & 21.91$\pm$0.14 & 0.10$\pm$0.03 & 0.55$\pm$0.17 & 0.38$\pm$0.03 \\
F480M & 19.4\%$\pm$3.3\% & 22.96$\pm$0.20 & 21.40$\pm$0.04 & 0.16$\pm$0.03 & 0.91$\pm$0.16 & 0.48$\pm$0.03 \\
\hline
\end{tabular}
\tablecomments{The values presented in this table are derived from the host galaxy fitting for J2236+0032 in eight bands. These results are based on the weighted result using different fitting configurations as introduced in Section~\ref{weighting_stra}. The corresponding decomposition images are presented in Figure E1a in~\cite{Onoue2024}.}
\label{table:J2236_result}
\end{table*}

\subsection{Host stellar masses from SED fitting}
\label{subsec:sed}
We estimate the stellar masses of the host galaxies using spectral energy distribution (SED) fitting, following a similar methodology as that introduced by~\citet{Ding2023}. For most targets, we adopt the photometric data from the NIRCam F356W and F150W filters, which bracket the rest-frame Balmer break at the redshifts of our sample. For objects with non-significant detections in F150W, we apply the flux upper limits from Table~\ref{tab:F150W_result} to constrain the model; for J1146$-$0005, the one object with a non-significant detection in F356W, we incorporate the value of $1\text{-}\sigma$ flux limits as the data points in the SED fitting and adopt the corresponding \smass\ as an upper limit. A special case is J2236+0032, for which we adopt the host fluxes derived from all eight NIRCam filters (Section~\ref{sec:J2236}) to achieve a more robust SED fit.

To infer the host stellar mass, we assume a~\cite{Chabrier2003} IMF for our stellar population model, which is consistent with the control sample from COSMOS-Web.\footnote{The \cite{Chabrier2003} IMF yields stellar mass estimates consistent with the \cite{Kroupa2001MNRAS.322..231K} IMF (differences $<0.05$~dex) but produces masses $\sim0.1\text{-}0.25$ dex smaller than those derived using the \cite{Salpeter1955ApJ...121..161S} IMF.} Our SED templates are governed by three key parameters: stellar population age, metallicity (Z), and dust attenuation (A$\mathrm{_{V}}$, rest-frame V-band extinction)\footnote{As shown by~\citet[Extended Data Figure 5]{Ding2023}, the stellar mass exhibits limited sensitivity to age and metallicity but is significantly more sensitive to A$\mathrm{_{V}}$.}.
Dust attenuation, which exhibits the strongest degeneracy with stellar mass, is assigned a prior range of A$\mathrm{_{V}}$=[0, 5.0] mag.  For host galaxies undetected in the F150W filters, the limited constraining power of the photometry motivates the use of a log-normal prior with a median value of 0.85, consistent with recent JWST observations of high-redshift galaxies~\citep{Shapley2023ApJ...954..157S}. Our approach incorporates a broad range of A$\mathrm{_{V}}$ values, which enhances the likelihood of encompassing the true values of the stellar mass. In contrast, the stellar population age has less impact on the inferred mass estimates. We adopt a fiducial uniform prior of [0.01, 0.84] Gyr, extending to the age of the Universe at $z\sim$6.5\footnote{If instead, we fix the age to a typical value for a $z\sim6$ galaxy, such as 0.3 Gyr, the inferred stellar mass changes by $\lesssim0.2$ dex for all our sources.}. We find the metallicity is even less influential: fixing Z/Z$_{\rm \odot} = 0.5$ (adopted in this work) results in $<0.1$~dex difference compared to fixing Z/Z$_{\rm \odot} = 0.1$.
To account for nebular contributions, we include a uniform prior on the ionization parameter ($\log$ U) over the range [$-3$, $-1$]. The robustness of our stellar mass measurements also stems from having photometric data on both sides of the 4000 \AA~break, accurate spectroscopic redshifts, and a firmly constrained upper limit on stellar age (0.84 Gyr), considering the age of the Universe at $z\sim6.4$.

We perform SED model fitting using the {\sc gsf} software~\citep{Morishita2019}, which generates a range of stellar templates with various ages and metallicities according to the defined priors to fit a composite stellar population-like star formation history. Parameter inference is performed via Markov Chain Monte Carlo (MCMC) sampling to derive the probability distribution of the SED parameters. The {\sc gsf} software was tested in \cite{Ding2023} by comparing its SED results with those obtained from other independent codes, such as {\sc bagpipes}~\citep{Carnall2018} and {\sc cigale}~\citep{Boquien2019}, for SED fitting; these comparison tests show minimal variations, with general differences in the derived stellar mass (\smass) values being 0.1 dex or less.

Given the limited ability of our two-band photometry in constraining additional SED parameters (such as age and metallicity), we focus exclusively on presenting the inferred host stellar mass in this work. We present the final inferred host stellar masses in Table~\ref{tab:table4} and show inferred two-band SED results in Figure~\ref{fig:SED_fitting}.

\begin{figure*}
\centering
\begin{tabular}{c c}
{\includegraphics[trim = 0mm 0mm 0mm 0mm, clip, height=0.3\textwidth]{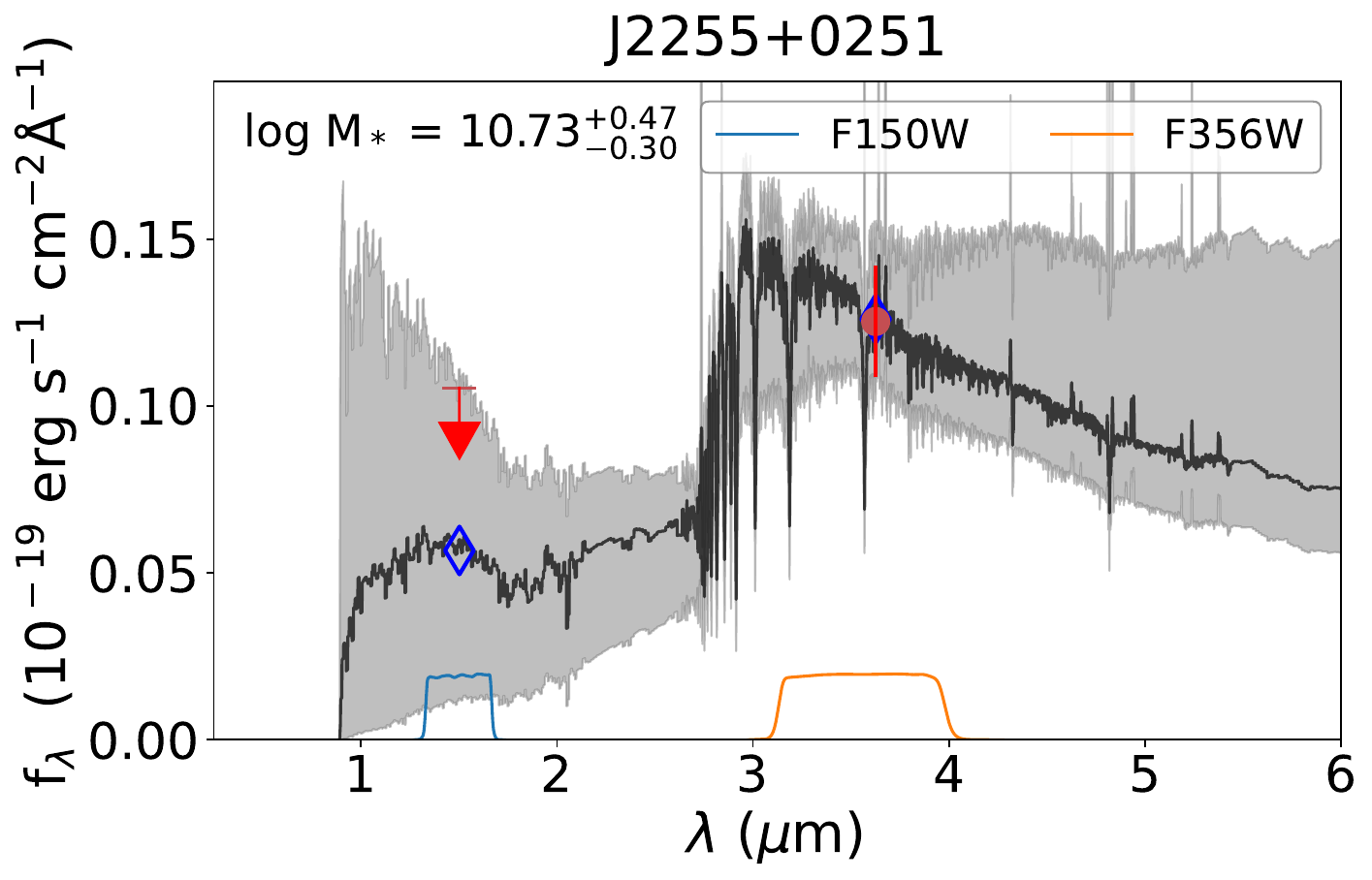}}&
{\includegraphics[height=0.3\textwidth]{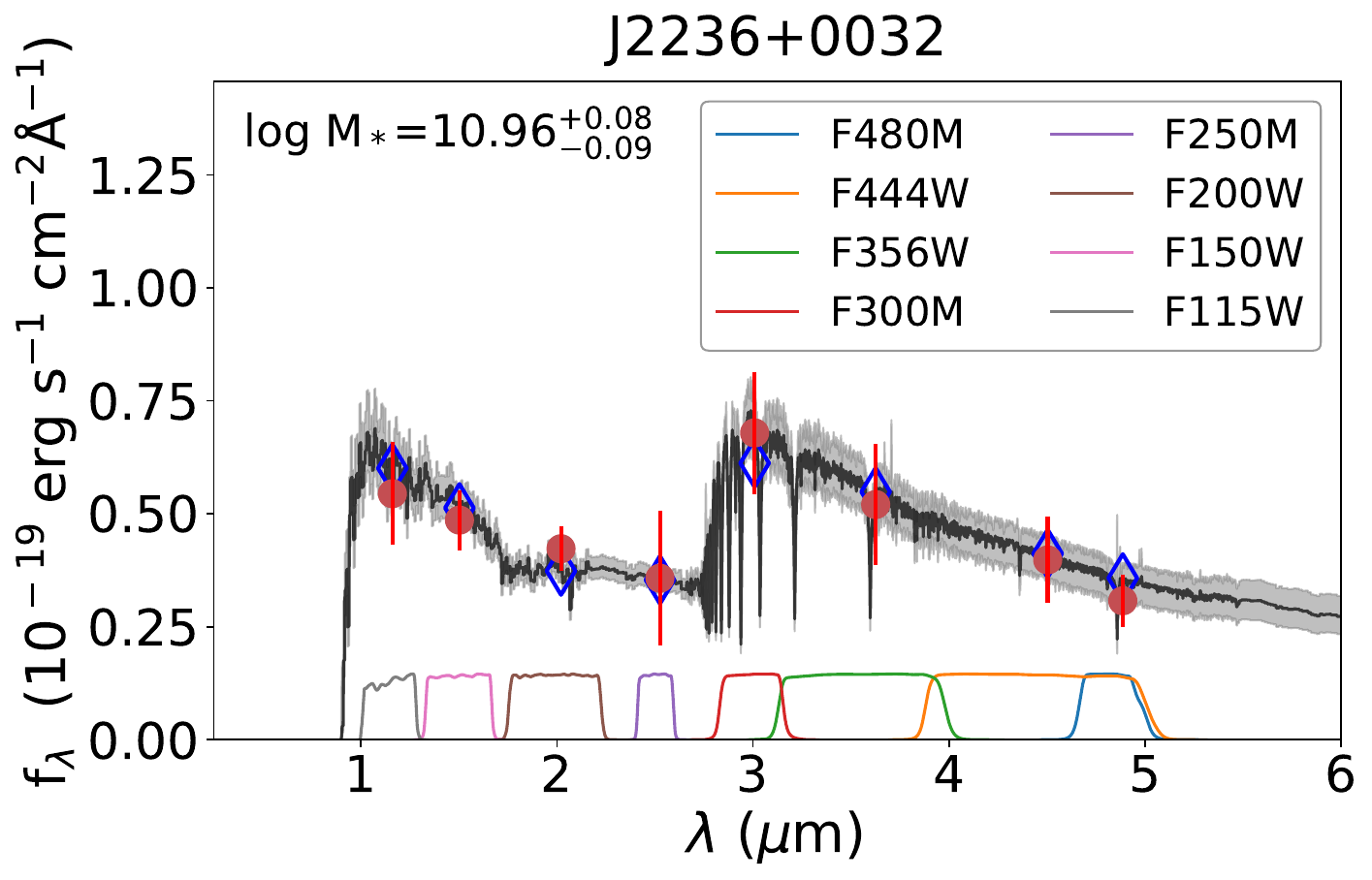}}\\
{\includegraphics[trim = 0mm 0mm 0mm 0mm, clip, height=0.3\textwidth]{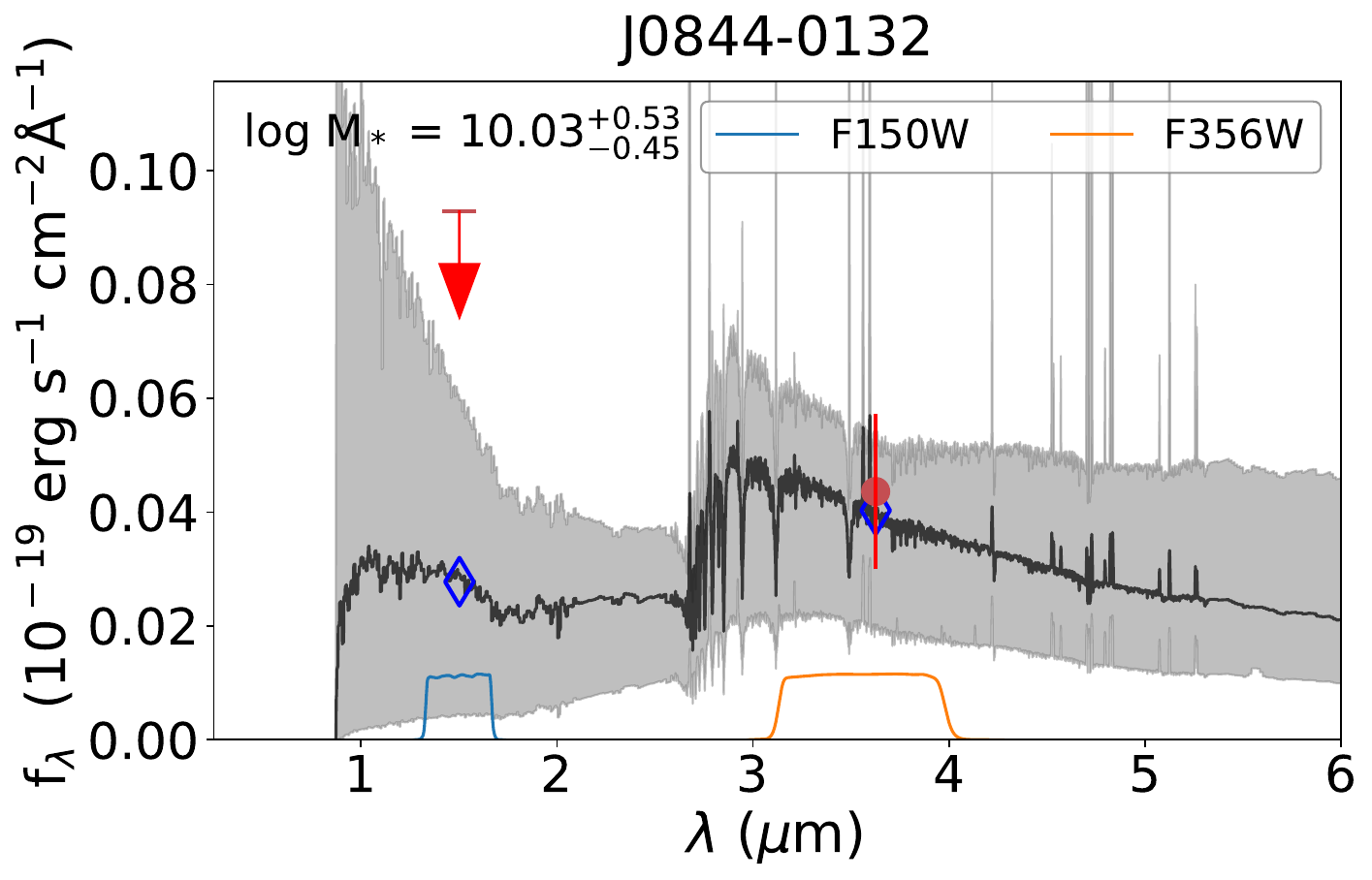}}&
{\includegraphics[height=0.3\textwidth]{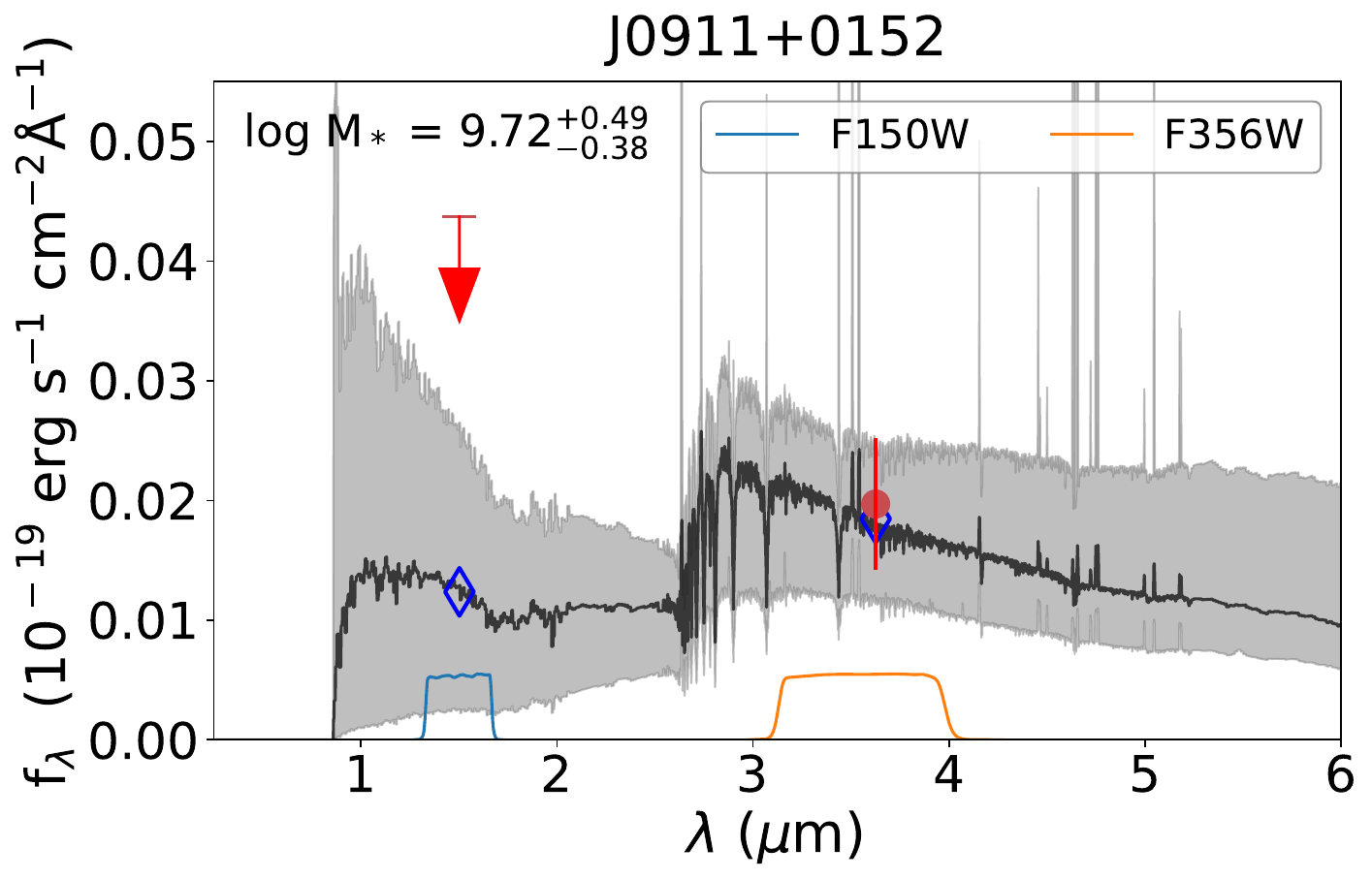}}\\
{\includegraphics[trim = 0mm 0mm 0mm 0mm, clip, height=0.3\textwidth]{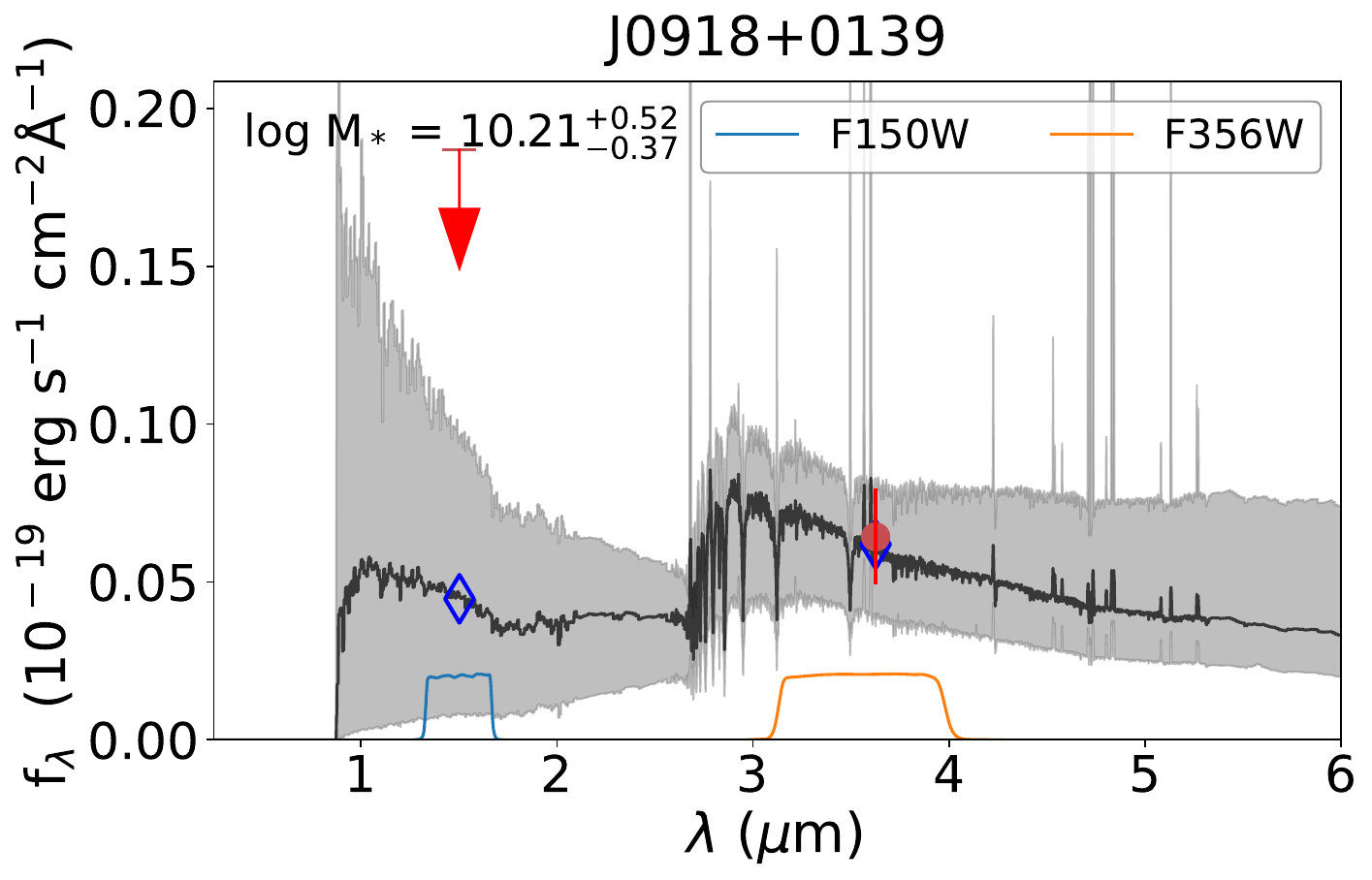}}&
{\includegraphics[height=0.3\textwidth]{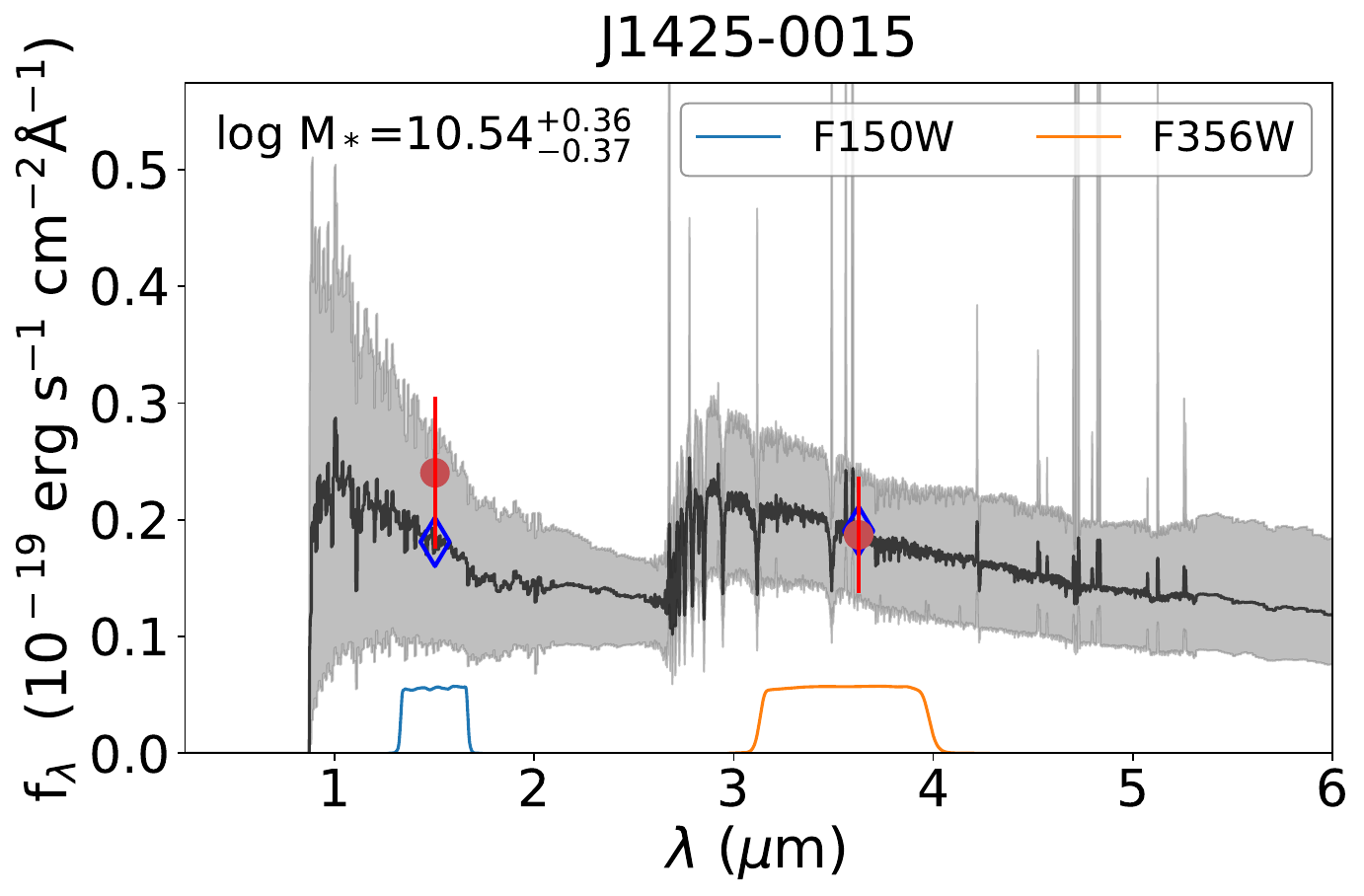}}\\
\end{tabular}
\caption{This figure presents the SED inference for our host galaxies, derived using the {\sc gsf} software. Red data points with error bars represent the inferred host galaxy fluxes across multiple bands. The gray-shaded region illustrates the 1-$\sigma$ range of SED template variations obtained through MCMC sampling. The black line denotes the median SED template. Blue diamonds indicate the flux predictions based on this median template for each observed band. We use eight-band photometry to infer the SED for J2236+0032~\citep{Onoue2024}.
\label{fig:SED_fitting} }
\end{figure*} 

\setcounter{figure}{4}  
\begin{figure*}
\centering
\begin{tabular}{c c}
{\includegraphics[trim = 0mm 0mm 0mm 0mm, clip, height=0.3\textwidth]{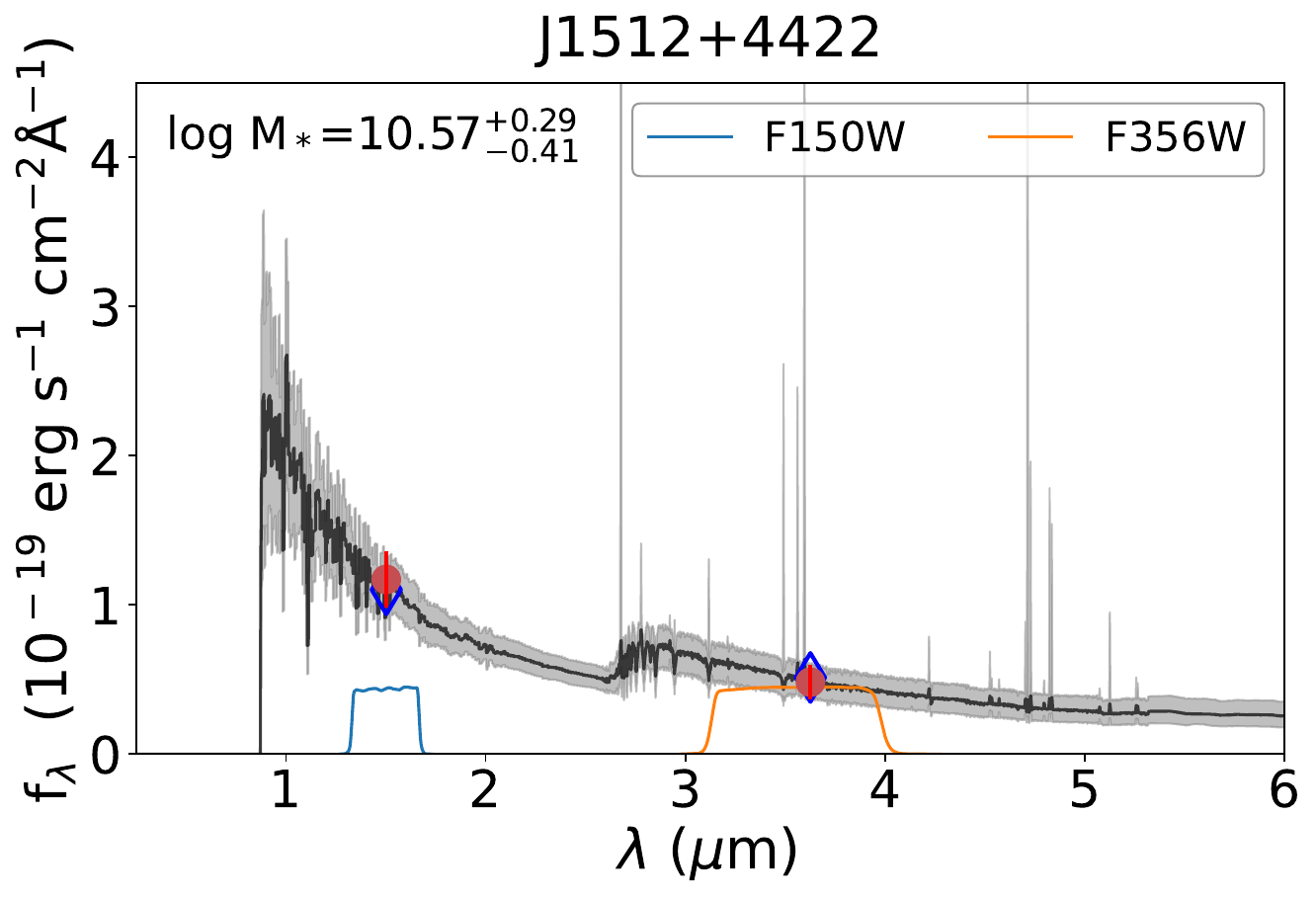}}&
{\includegraphics[height=0.3\textwidth]{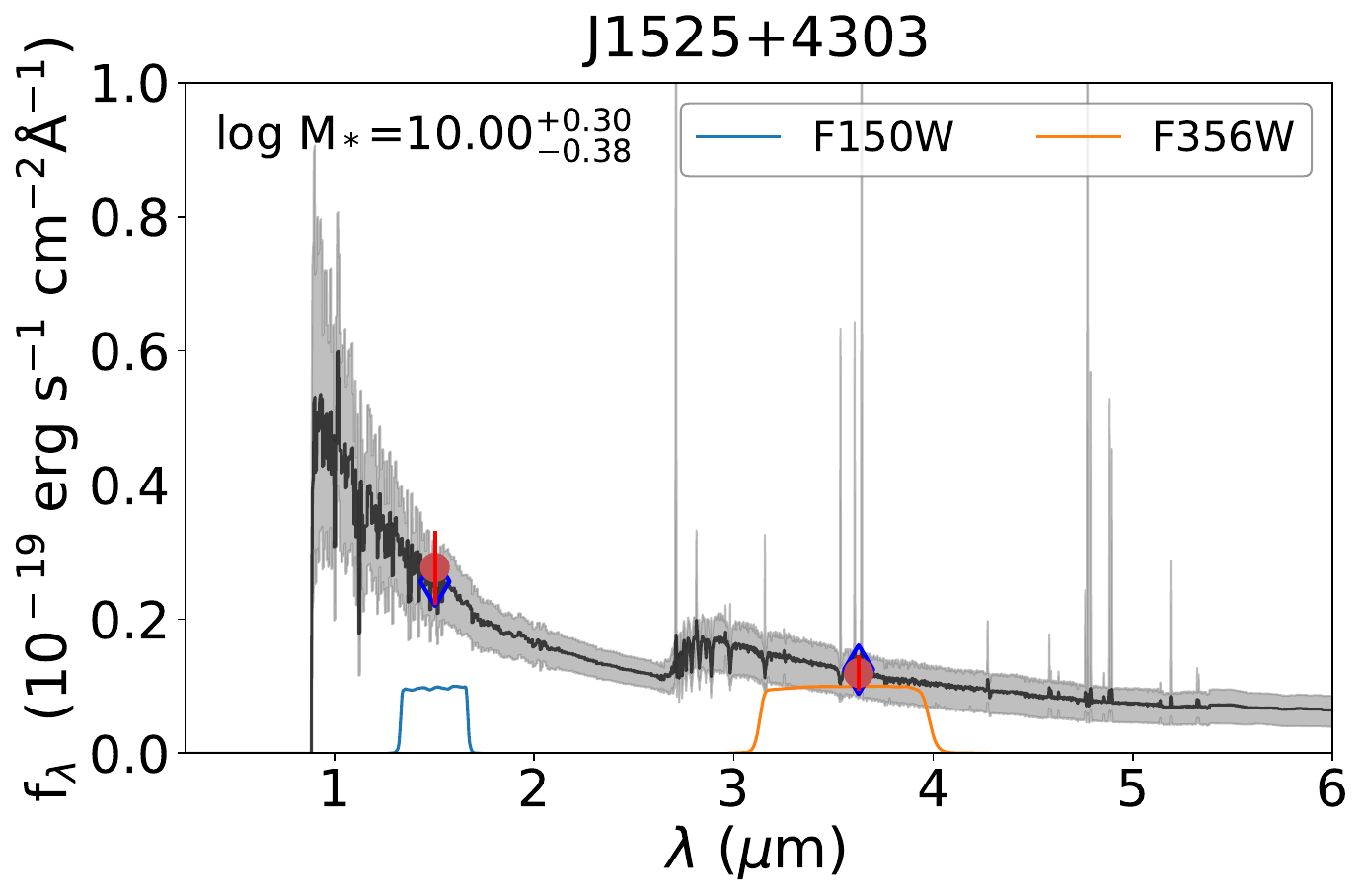}}\\
{\includegraphics[trim = 0mm 0mm 0mm 0mm, clip, height=0.3\textwidth]{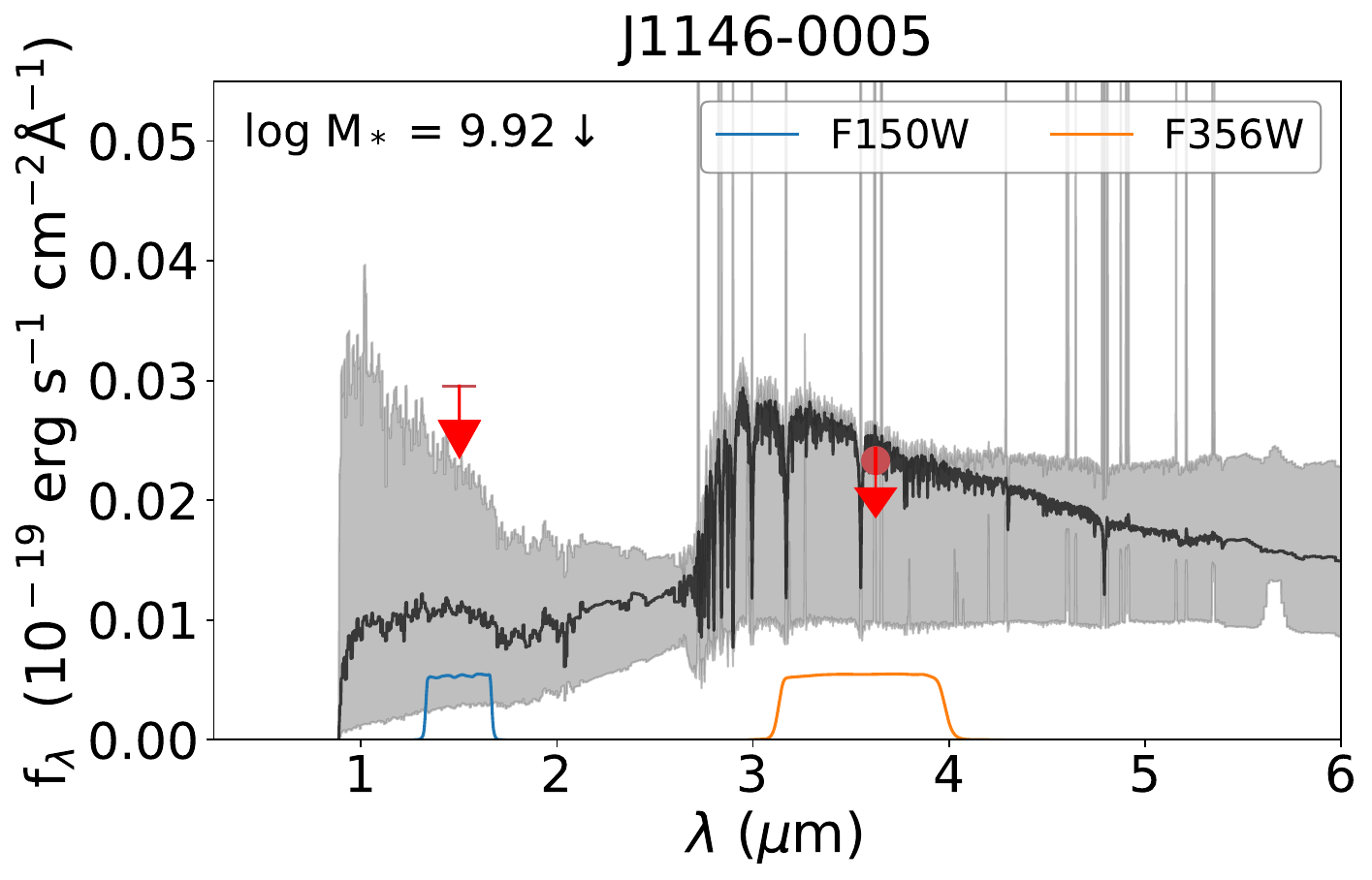}}&
{\includegraphics[height=0.3\textwidth]{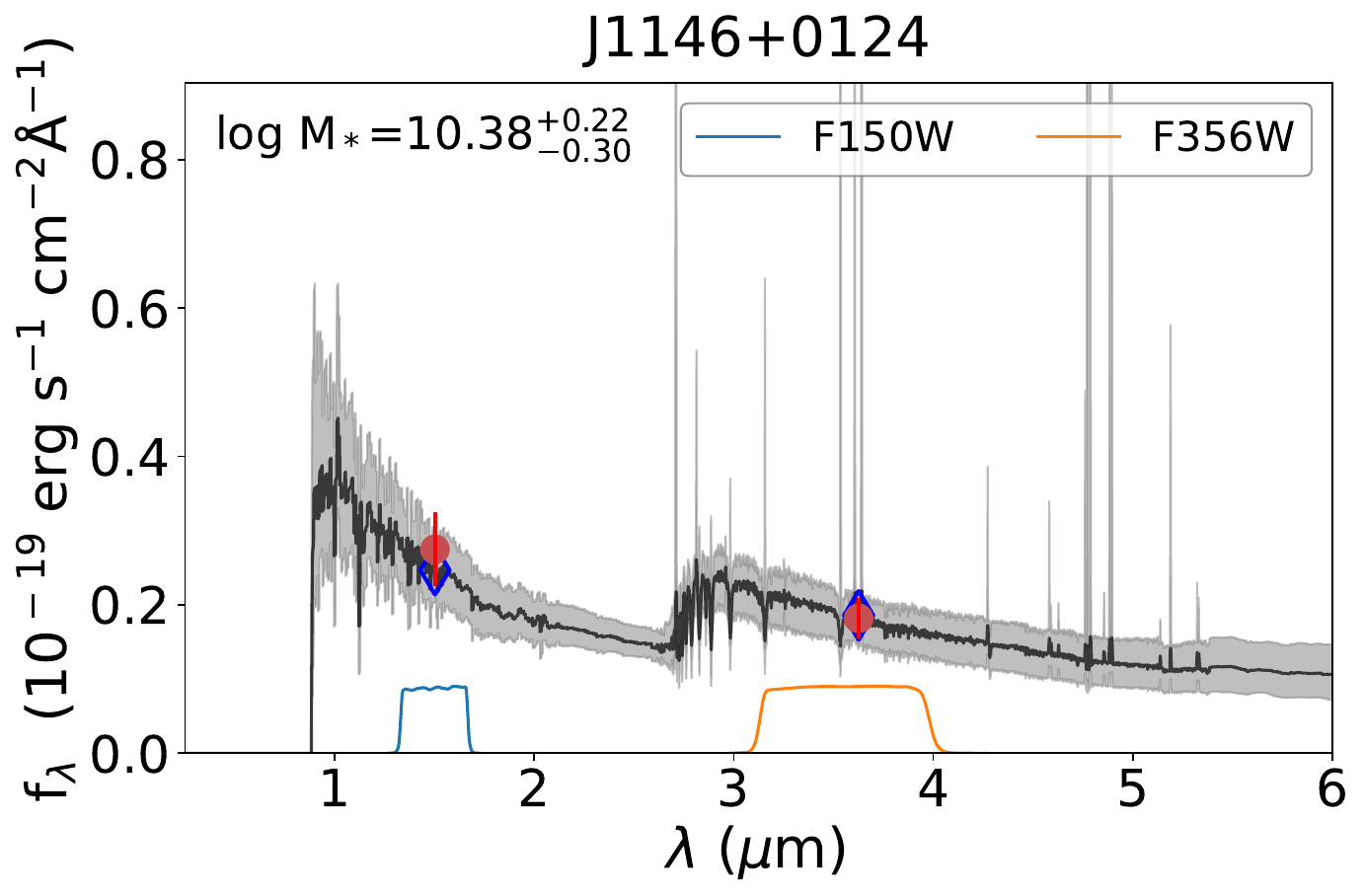}}\\
{\includegraphics[trim = 0mm 0mm 0mm 0mm, clip, height=0.3\textwidth]{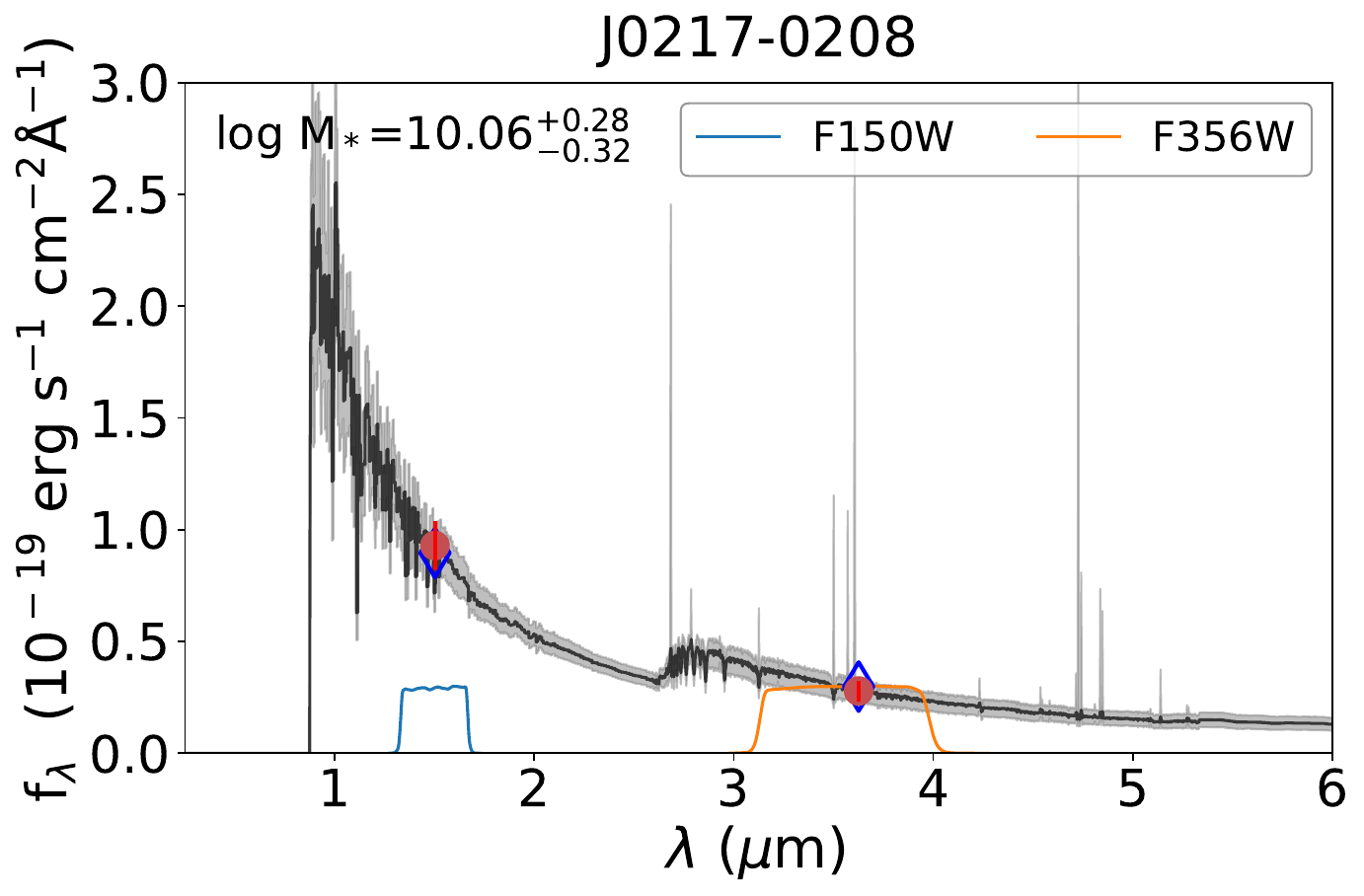}}&
{\includegraphics[height=0.3\textwidth]{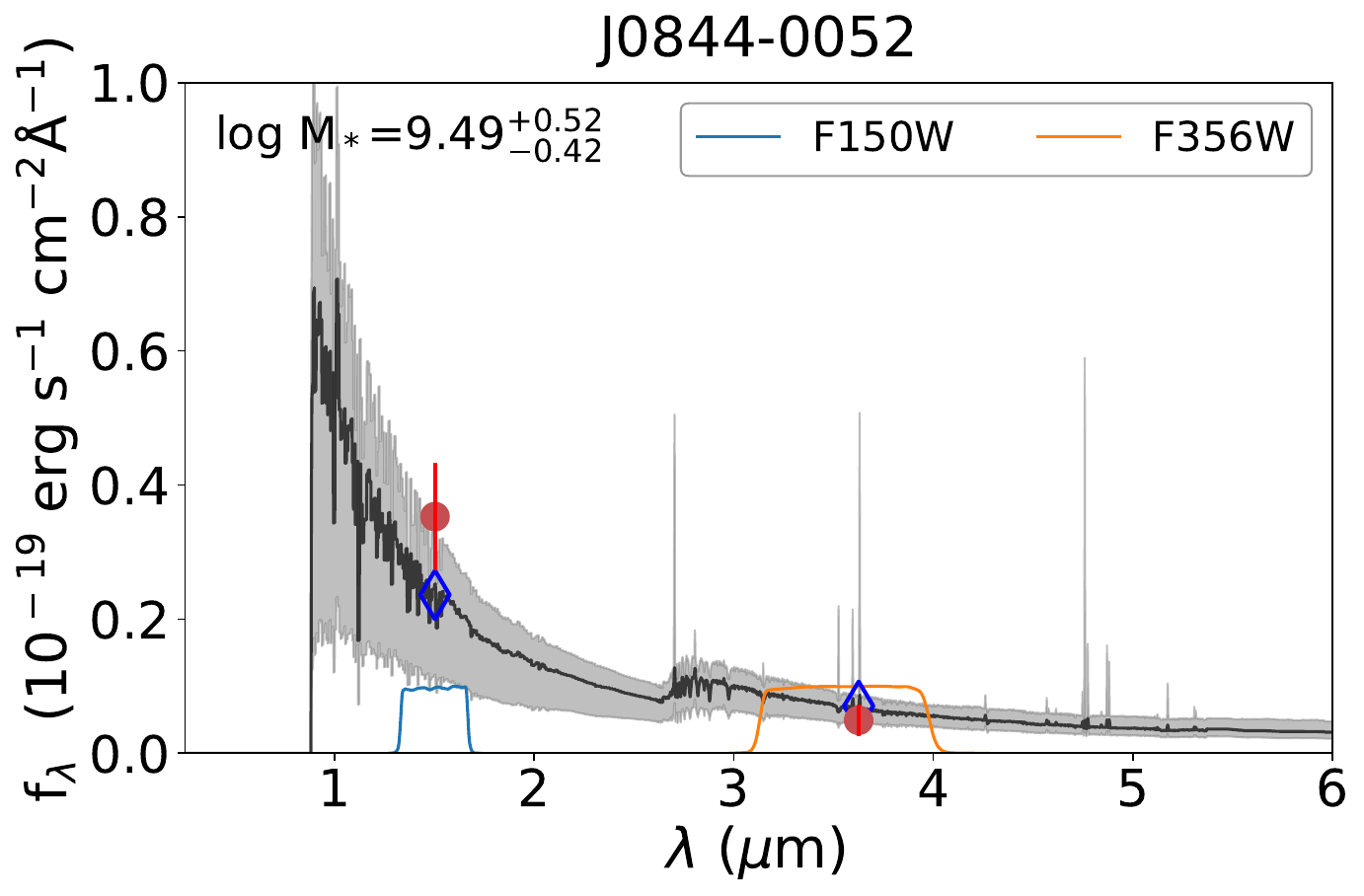}}\\
\end{tabular}
\caption{Continued. 
For J1146$-$0005, the inferred host detection in F356W is weak and does not pass the detection criteria. Thus, the inferred stellar mass for this target should also be considered as an upper limit.}
\end{figure*} 

The stellar masses of the quasar host galaxies span a wide range, from $10^{9.5}-10^{11.0}$~M$_{\rm \odot}$, with a median value of $10^{10.5}$~M$_{\rm \odot}$. The general systematic uncertainty of the stellar mass is 0.35 dex. Notably, the host galaxies of J2255+0251 have stellar masses of ${10.57}_{-0.41}^{+0.29}$ (in units of log~M$_{*}$/M$_{\rm \odot}$), consistent with our previous studies~\citep[i.e., ${10.53}_{-0.37}^{+0.51}$, ][]{Ding2023}. 
For J2236+0032, the fitting result (${10.96}_{-0.09}^{+0.08}$) is based on eight-band photometry; the inferred \smass\ is also consistent with \cite{Ding2023} (${11.12}_{-0.27}^{+0.40}$) based on two-band fitting (F150W+F356W). This result is also consistent with the reported value in \cite{Onoue2024} ($10.81\pm{0.08}$) in which additional information from JWST/NIRSpec spectroscopy is applied during SED fitting.

On the other hand, less massive hosts, such as J0911+0152, J0844$-$0052, and J1146$-$0005, exhibit stellar masses of log~M$_{*}$/M$_{\rm \odot} < 9.9$, respectively, with the latter being a $1\text{-}\sigma$ upper limit due to lack of detection in the F356W filter. The diversity in stellar masses reflects the varying evolutionary stages and accretion histories of these high-redshift quasar hosts. Importantly, the stellar masses of our sample are consistent with those of massive star-forming galaxies at similar redshifts as reported in \cite{Yang2025arXiv250407185Y}, suggesting that these quasar hosts are among the most massive galaxies in the early universe (see next subsection).

\begin{table}
    \centering
    \caption{Inferred stellar mass of the quasar hosts}
    \begin{tabular}{cc|cc}
\hline\hline
ID &  log M$_{*}$/M$_{\rm \odot}$ & ID &  log M$_{*}$/M$_{\rm \odot}$ \\
\hline
J2255+0251 & ${10.73}_{-0.30}^{+0.47}$ & 
J2236+0032 & ${10.96}_{-0.09}^{+0.08}$ \\ 
J0844$-$0132 & ${10.03}_{-0.45}^{+0.53}$ & 
J0911+0152 & ${9.72}_{-0.38}^{+0.49}$ \\ 
J0918+0139 & ${10.21}_{-0.37}^{+0.52}$ & 
J1425$-$0015 & ${10.54}_{-0.37}^{+0.36}$ \\ 
J1512+4422 & ${10.57}_{-0.41}^{+0.29}$ & 
J1525+4303 & ${10.00}_{-0.38}^{+0.30}$ \\ 
J1146$-$0005 & $9.92\downarrow$ & 
J1146+0124 & ${10.38}_{-0.30}^{+0.22}$ \\ 
J0217$-$0208 & ${10.06}_{-0.32}^{+0.28}$ & 
J0844$-$0052 & ${9.49}_{-0.42}^{+0.52}$ \\ 
       \hline
    \end{tabular}
    \tablecomments{The inferred stellar mass based on the SED fitting approach introduced in Section~\ref{subsec:sed}. The $1\text{-}\sigma$ upper limit stellar mass (i.e., non-significant detection in F356W) is indicated with the downarrow symbol.
    The object J2236+0032 is constrained using eight bands and thus has a smaller uncertainty. The stellar mass values for J2236+0032 and J1512+4422 are also consistent with the inference in~\cite{Onoue2024} (i.e., $10.81\pm{0.08}$ and $10.63\pm{0.02}$, respectively), in which data from JWST/NIRSpec was applied in the SED fits. 
    }
    \label{tab:table4}
\end{table}

\begin{figure*}
\centering
\begin{tabular}{c c}
{\includegraphics[trim = 0mm 0mm 0mm 0mm, clip, height=0.45\textwidth]{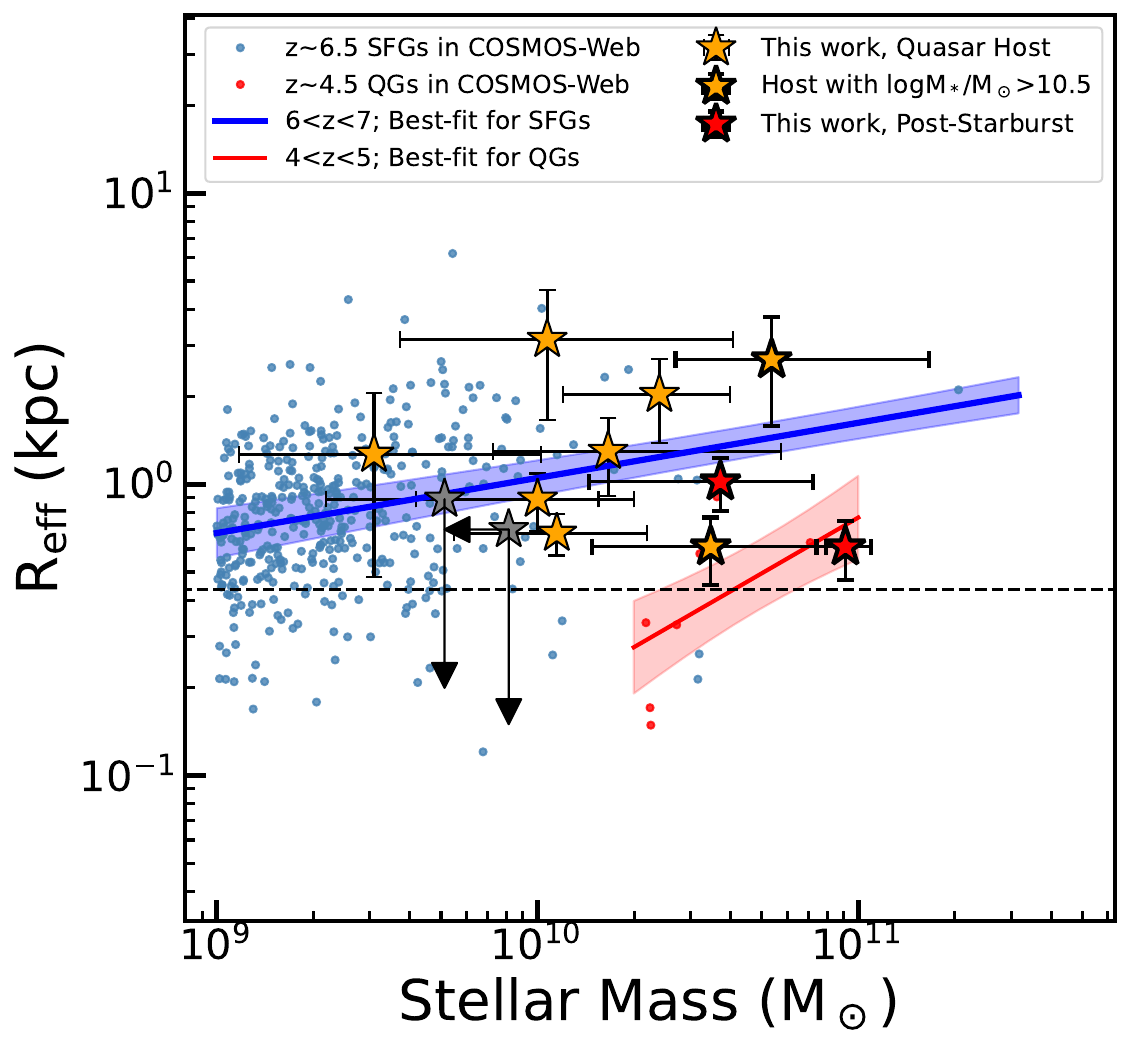}}&
{\includegraphics[height=0.45\textwidth]{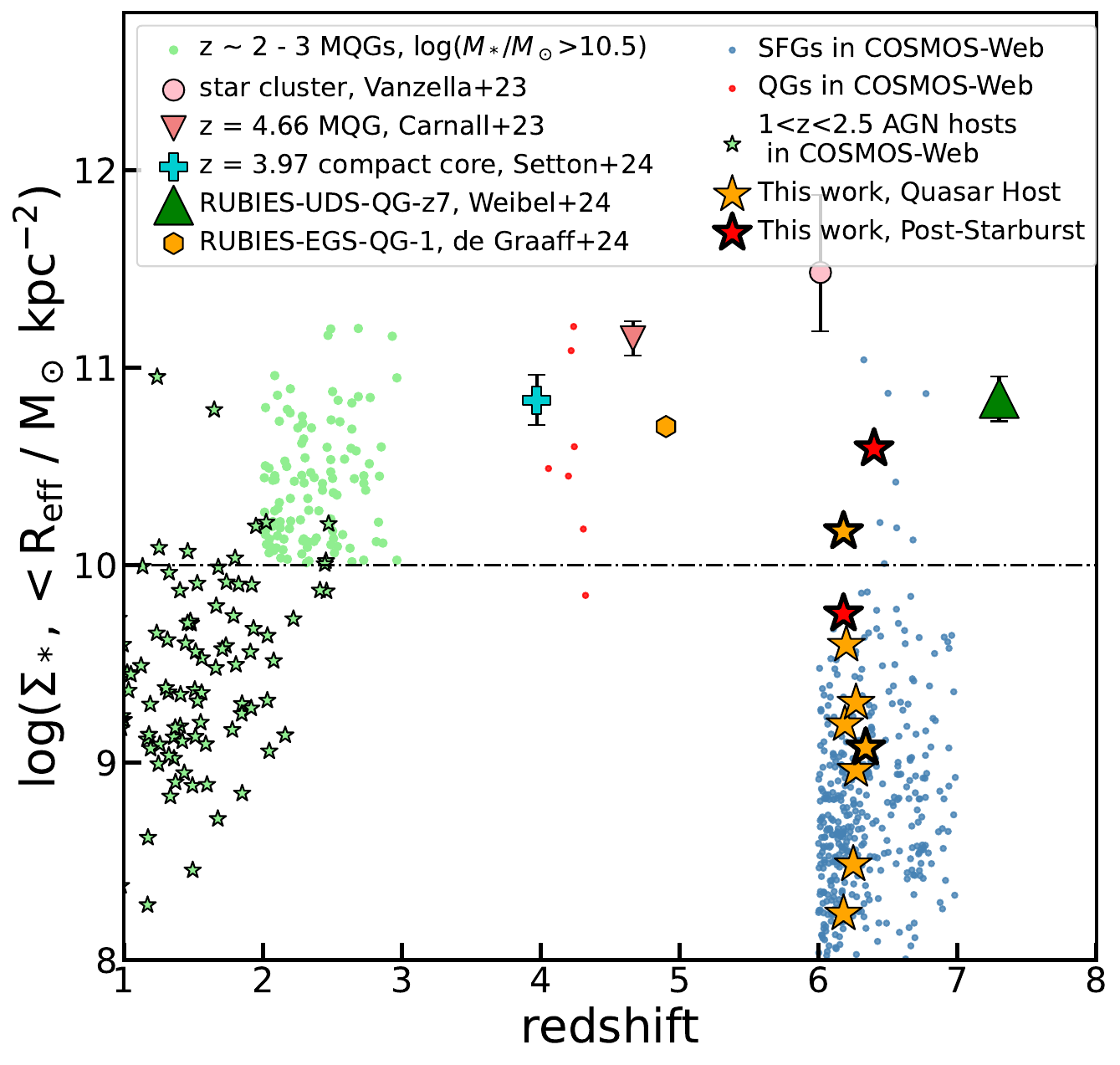}}\\
\end{tabular}
\caption{\label{fig:size-mass}
(Left) Size - stellar mass relation of our quasar host galaxies compared with the control sample of star-forming galaxies (SFGs) and from COSMOS-Web \citep[see][]{Yang2025arXiv250407185Y} at redshift $6<z<7$. The two gray stars with arrows are the host measurements with size or stellar mass as upper limits. We also include seven quiescent galaxies (QGs) at $4<z<5$ for comparison. In addition, the best-fit relations for the COSMOS-Web sample, derived assuming the power-law model, are also shown.
We measure the sizes of our sample host galaxies using the F356W filter, while the COSMOS-Web sample adopts the F444W filter (see Section~\ref{subsec:SizeMass} for methodological details).
The results show that the sizes of our quasar hosts appear to be similar to the SFGs for a given stellar mass, while the two post-starbursts are consistent with the QGs at lower redshifts within the 1-$\sigma$ level. The dashed horizontal line indicates the PSF size at redshift 6.
(Right) Projected 2D stellar mass surface density within \reff\ as a function of redshift, compared with those of quiescent galaxies from the literature~\citep{Skelton2014, Carnall2023, Vanzella2023ApJ, de_Graaff2024, Setton2024, Weibel2024} as well as with $1<z<2.5$ AGN host galaxies revealed by COSMOS-web~\citep{Tanaka2024}.
Most of our quasar hosts fall below the typical quiescent threshold~\citep[dashed line of $10^{10}$~M$_{\rm \odot}$/kpc$^2$ as introduced in][]{Weibel2024}. The two post-starburst targets are in the top region of our sample. The typical uncertainty associated with the density measurements of the hosts in our sample is 0.5~dex.}
\end{figure*}

\subsection{Size--mass relation}\label{subsec:SizeMass}
We investigate the stellar size--mass relation of our 12 quasar host galaxies and present their distribution in Figure~\ref{fig:size-mass}~(left). The sizes for our sample range from 0.5 to 3 kpc (represented by the star symbols), indicating a wide variation at a given stellar mass. For comparison, we show the size measurements of a control sample of non-active star-forming galaxies at $z\sim6.5$ from the COSMOS-Web survey~\citep{Yang2025arXiv250407185Y}, along with the best-fit relation. Instead of applying an empirical formula for a wavelength correction on the sizes, we directly use the measured values in the long wavelength filter: F356W for our quasar host sample and F444W for the COSMOS-Web sample. While these sizes are derived from two different filters, we expect their apparent sizes to be similar since both bands are above 4000~\AA~and are in the rest-frame optical regime. This expectation is further supported by the multi-band decomposition results for J2236+0032 (see Table~\ref{table:J2236_result}). 
The size--mass relationship for our objects is close to that of the $z\sim6.5$ star-forming sample. That is, our quasar hosts have sizes comparable to those of COSMOS-Web star-forming galaxies with the same stellar mass.

Our control sample from the COSMOS-Web survey at $z>6$ includes very few quiescent galaxies, limiting direct comparisons within this redshift range. To address this population, we use a sample of seven quiescent galaxies at $z\sim4\text{--}5$ to facilitate our understanding of evolutionary trends; the results indicate that the sizes of our quasar hosts are generally larger than the distribution for lower-redshift quiescent galaxies. Given that galaxy sizes of the COSMOS-Web sample are expected to be more compact with increasing redshift, our result implies that quasar hosts at $z>6$ should be larger than quiescent galaxies of comparable stellar masses at similar redshifts. Notably, the two post-starburst quasar hosts in our sample, J2236+0032 and J1512+4422 \citep[reported by][]{Onoue2024}, exhibit sizes that are consistent at 1-$\sigma$ level with the lower-redshift quiescent galaxy distribution. This agreement suggests that their morphological properties are consistent with those of quiescent galaxies, which is expected given their post-starburst nature, meaning that they will likely transition toward quiescence.

We note that the F356W filter at $z\sim6$ includes the H$\beta$ and [O${\rm III}$] emission lines. If these nebular lines are spatially extended, they could, in principle, bias the measured host galaxy sizes toward larger values, since the ionized gas distribution can be more extended than the stellar continuum. However, as reported in our previous work based on JWST/NIRSpec spectroscopy~\citep[][Onoue et al. in prep., and Phillips et al. in prep.]{Ding2023, Onoue2024}, the H$\beta$ and [O${\rm III}$] emission lines are generally weak in our sample. For example, in J2236+0032, the narrow H$\beta$ emission is not detected, and the [O${\rm III}$] emission in NIRSpec contributes roughly $3\%$. This indicates that nebular line contamination is minimal, and the stellar continuum emission dominates the measured F356W sizes.

\subsection{Stellar mass density}

We estimate the 2D stellar mass surface density 
by assuming that half of the stellar mass is within \reff. We then compare it with the COSMOS-Web sample as well as quiescent galaxies from the literature across a broad redshift range, as shown in Figure~\ref{fig:size-mass}~(right). The quiescent galaxies included in this comparison consist of those at $z\sim2\text{--}3$ from the 3DHST catalog \citep{Skelton2014}, along with a sample of massive quiescent galaxies at $z\gtrsim4$  \citep{Carnall2023, Vanzella2023ApJ, de_Graaff2024, Setton2024, Weibel2024}. Our analysis reveals that most of our quasar hosts fall below the empirical lower limit for quiescent galaxies in the literature (i.e., log($\Sigma_*$) $\sim$ 10). Notably, the two post-starburst galaxies, J2236+0032 and J1512+4422, occupy the upper region of this density space, aligning with the characteristics of quiescent galaxies.

\begin{figure*}
    \centering
    \includegraphics[width=0.8\linewidth]{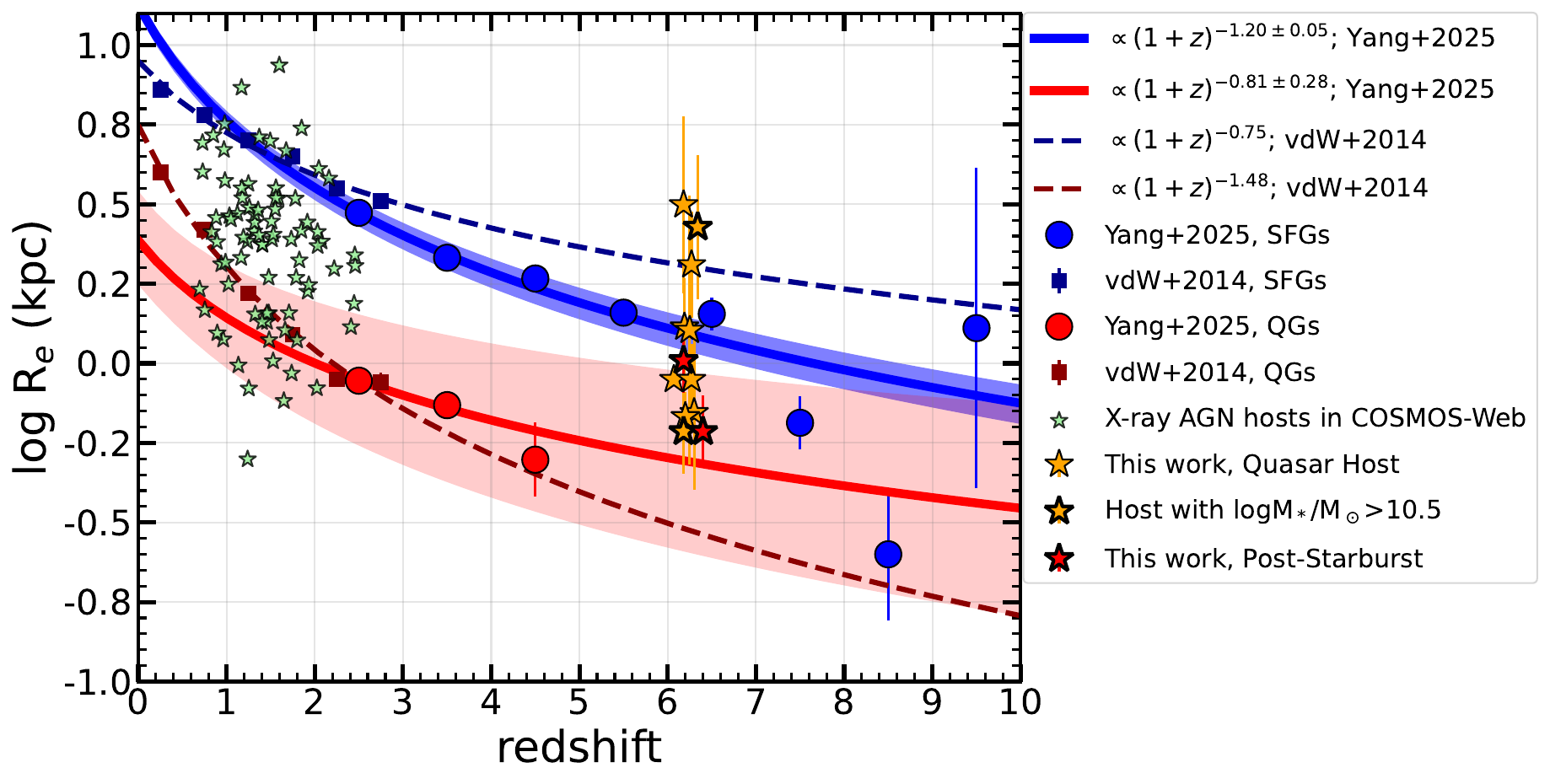}
    \caption{Size evolution of our quasar hosts with redshift, comparing them to the rest-frame optical sizes of star-forming and quiescent galaxies at a fixed stellar mass of 5$\times10^{10}$ M$_{\rm \odot}$ at $0 <z <10$; the comparative observational measurements and evolution trends are provided by \cite{van_der_Wel2014, Yang2025arXiv250407185Y}. We also incorporate recent quasar host size measurements from the COSMOS-Web survey, as reported by \cite{Tanaka2024}. The result suggests that most massive hosts (log~M$_{*}$/M$_{\rm \odot} >10.5$)  have systematically smaller effective radii, approaching the compact sizes characteristic of quiescent galaxies.
    }
    \label{fig:size_evo}
\end{figure*}

\subsection{Local environments of quasar hosts}
The local environments of our quasar hosts reveal a mix of isolated systems and those with nearby companion galaxies. Six of 12 targets, exhibit neighbors within a projected separation of 2\farcs1 in the F356W images, corresponding to a physical distance of 12 proper kpc at $z\sim6$. Notably, two quasars, J0844$-$0032 and J0217$-$0208 are accompanied by nearby objects with projected separations of less than 0\farcs3, which may indicate ongoing interactions or minor mergers. The companion galaxy near J1512+4422 has been identified as a foreground object based on our NIRSpec observations (Phillips et al. in prep). 
However, the majority of the quasar hosts appear relatively isolated, with no strong evidence of major mergers or significant tidal features in their immediate environments. This may reflect the limitations of current observations, including low host galaxy SNR and surface brightness sensitivity, rather than a true absence of ongoing mergers or interactions at these redshifts.

To further investigate the local environments of our quasar hosts, we compare the frequency of nearby neighbors in our quasar sample to that of star-forming galaxies in the COSMOS-Web survey at similar redshifts and similar stellar masses. In COSMOS-Web, approximately 60\% (68 out of 113) of star-forming galaxies have at least one apparent neighbor within the same projected separation level (i.e., 2\farcs1 $\times$ 2\farcs1) in the F444W filter at the similar stellar mass (log M$_{*}$/M$_{\rm \odot}>9.5$). This fraction is consistent with the 50\% (6 out of 12) observed in our quasar sample, suggesting that the local environments of quasar hosts are not significantly different from those of typical star-forming galaxies at $z\sim6$.
We note that this fraction is also consistent using the dusty companions as seen in ALMA observations using high-$z$ luminous quasars ~\citep[e.g.,][]{Decarli2017Natur.545..457D, Neeleman2021ApJ...911..141N, Meyer2022ApJ...927..141M}, which is $\sim30\%$.
We note that the current analysis focuses on companions within a 2\farcs1 projected separation; however, the larger field of view of our JWST observations allows for examining galaxy overdensities on broader scales. We plan to conduct a comprehensive study of the quasar environments as a function of scale in future work.

The lack of morphological signatures of recent mergers (e.g., tidal tails and asymmetric clumps) in 11/12 quasar hosts, with the exception of J0844$-$0132, challenges the traditional paradigm that major mergers are the dominant mechanism for triggering AGN activity at high redshifts.
Instead, our results align with emerging evidence that secular processes, such as disk instabilities or minor interactions, may play a more significant role in fueling AGNs during the epoch of reionization. This is consistent with recent studies of lower-redshift quasars, which have shown that AGN activity can be sustained in relatively isolated galaxies without the need for major mergers~\citep[e.g.,][]{Marian2019ApJ...882..141M}. However, the presence of close companions in some systems, such as J0844$-$0032 and J0217$-$0208, suggests that interactions may still contribute to AGN fueling in certain cases. These systems provide valuable opportunities to study the role of minor mergers and interactions in driving black hole growth and star formation in the early universe. Future follow-up spectroscopic observations (e.g., JWST-GO-7519, PI: Arita, using NIRCam WFSS observations) and ALMA observations~\cite[such as \text{[C\,{\sc ii}]} and CO,][]{Decarli2017Natur.545..457D, Decarli2019ApJ...880..157D}  will be crucial to confirm the redshifts of these companion galaxies and determine whether they are physically associated with the quasar hosts.


\section{Discussion} \label{sec:diss}

\subsection{Moderate-luminosity quasars as critical probes}\label{diss:faint_qso}
Moderate-luminosity quasars ($M_{1450}>-24$ mag) offer advantages for advancing our understanding of AGN-host galaxy coevolution in the following ways. First, their lower intrinsic luminosity ensures a higher host flux ratio (see Figure~\ref{fig:ratio}, top), thus increasing the observational success rate in terms of detecting and characterizing their host galaxies. In this work, we achieved host detections in 11 out of 12 targets (92\%), perhaps due to less contamination from nuclear emission than we would have at higher luminosities. Second, moderate-luminosity quasars dominate the AGN population by number, representing the most statistically representative subset of the entire quasar distribution. This minimizes selection biases inherent in studies of rare, luminous systems \citep[e.g., the ``tip-of-the-iceberg" effect; see the discussion in][]{Li2025ApJ...981...19L}, which often overrepresent extreme accretion states or environments. Our findings highlight the need to study moderate-luminosity quasars as analyzed in this work to establish a robust and unbiased picture of black hole growth and its connection to host galaxy evolution across cosmic time.

\subsection{Implications for galaxy compaction models}\label{diss:compaction}
A key goal of this work is to investigate the structural properties of our $z>6$ quasar hosts within the framework of galaxy evolution. To this end, Figure~\ref{fig:size_evo} compares the physical size of our sample quasar hosts as observed in the rest-frame optical regime with the redshift evolution of star-forming and quiescent galaxies at a fixed stellar mass (\smass=5$\times10^{10}$ M$_{\rm \odot}$) from the observations by~\cite{Yang2025arXiv250407185Y} and \cite{van_der_Wel2014}. While the majority of our quasar hosts align with the size distribution of star-forming galaxies at $z\sim6$, we observe a striking trend: the most massive hosts (log~M$_{*}$/M$_{\rm \odot} >10.5$) exhibit systematically smaller effective radii, approaching the compact sizes of quiescent galaxies. 

This trend supports scenarios where massive galaxies undergo rapid structural transformation via compaction. In such models, gas-rich disks experience turbulent inflows driven by mergers, instabilities, or AGN feedback, triggering centrally concentrated starbursts that build dense cores \citep[e.g.,][]{Dekel2009ApJ...703..785D, Zolotov2015MNRAS.450.2327Z}. Subsequent quenching then preserves these compact morphologies, aligning them with the quiescent population. The two post-starburst quasar hosts in our sample reside at the high-mass end and show compact sizes with higher mass surface density, which is consistent with high-$z$ quiescent galaxies (i.e., Figure~\ref{fig:size-mass}). This consistency provides direct evidence for this sequence. Their properties suggest a phase where star formation is rapidly suppressed after a compaction-driven starburst, potentially linked to AGN activity.

Our results extend the validity of compaction models into the reionization epoch. The coexistence of extended, star-forming hosts and compact, massive systems in our $z>6$ quasar host sample implies that compaction is already underway in the earliest massive galaxies, with AGN activity potentially acting as both a catalyst (via feedback-driven inflows) and a consequence (via black hole growth during starbursts). In particular, feedback-driven inflows associated with AGN represent a key process in the evolution of massive galaxies, especially at high redshift. These inflows occur when energy and momentum from the central AGN, via outflows such as winds, jets, or radiation pressure, not only expel gas from the center but, under certain conditions, also induce the return of gas toward the nucleus. This phenomenon regulates both star formation and black hole growth~\citep[e.g.,][]{Zolotov2015MNRAS.450.2327Z}.

Future studies with dynamical mass measurements and spatially resolved star formation histories will test whether the compact hosts in our sample truly represent nascent quiescent cores, or instead are experiencing temporary structural changes that occur before star formation is fully suppressed.

\subsection{Systematics in host measurement}
\label{sec:systematic}
In this study, we employ state-of-the-art fitting techniques to measure the properties of quasar host galaxies. Utilizing a weighting algorithm, we combine results from various fitting settings (i.e., using different PSFs, changing PSF supersampling factor, and varying the value of the \sersic\ index $n$ from 1 to 4) to derive our final estimates. The final small scatter among the different inferences indicates a high fidelity in our host galaxy measurements, reinforcing the reliability of our findings.

\begin{figure*}
\centering
\begin{tabular}{c c}
{\includegraphics[trim = 0mm 0mm 70mm 0mm, clip, height=0.3\textwidth]{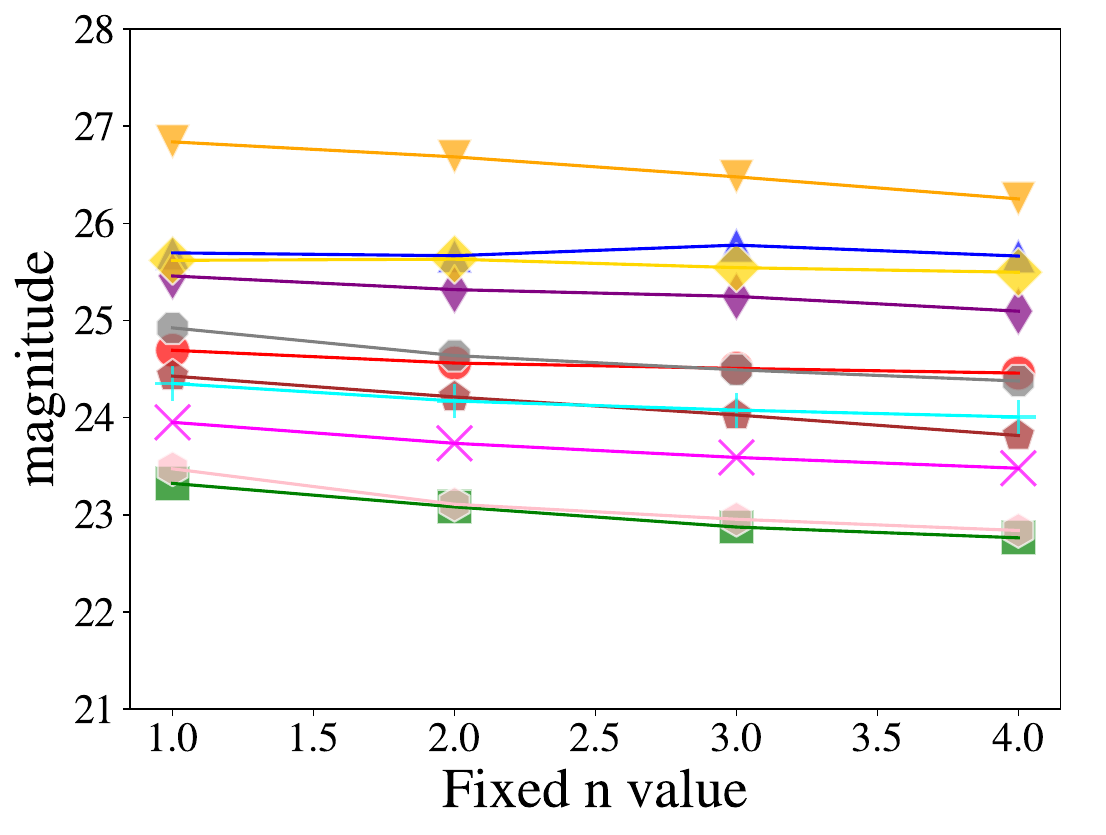}}&
{\includegraphics[height=0.3\textwidth]{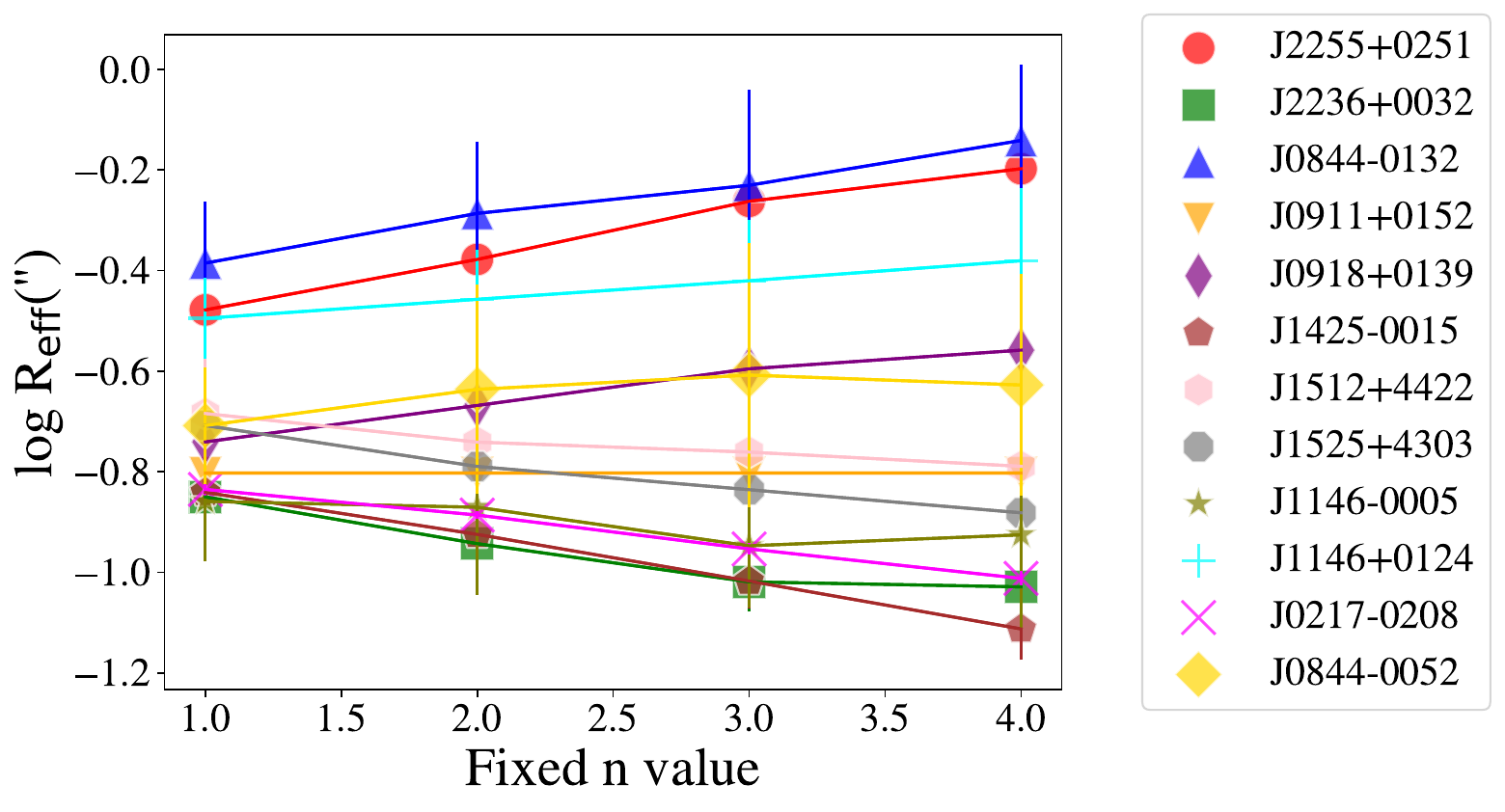}}\\
\end{tabular}
\caption{\label{fig:cy1_diff_n} 
This figure illustrates how the inferred host properties in NIRCam/F356W vary with changes in the fixed \sersic\ index $n$ values, ranging from 1 to 4. Since the \sersic\ index controls the central flux concentration, increasing $n$ results in a brighter inferred host. However, the response of the inferred effective radius (\reff) to changes in $n$ is not uniform. Generally, our results indicate a negative trend when the host is bright (mag $\lesssim24.5$) and compact (log \reff ($''$) $\lesssim-0.7$), while a positive trend emerges when the host is faint and extended. In this work, we adopt the combined results obtained by fixing the \sersic\ index to $n = 1, 2, 3,$ and $4$.
}
\end{figure*}

Figure~\ref{fig:cy1_diff_n} illustrates the variations in inferred host magnitude and \reff\ derived from the F356W filter as a function of the fixed \sersic\ index $n$. 
As shown, the inferred magnitudes are fairly insensitive to changes in the \sersic\ index, demonstrating that the final inferred stellar masses remain robust across a range of light profile shapes (i.e., from disk-like to bulge-like).
Our analysis reveals clear trends in how these parameters depend on different fixed $n$ values. We find that increasing the \sersic\ index $n$ leads to a brighter inferred host, which is expected since the index controls the concentration of light; higher values allocate more light from the total to the host galaxy~\citep{Treu2007ApJ...667..117T}. Interestingly, the response of \reff\ to $n$ differs: a negative trend is observed for bright hosts (mag $\lesssim 24.5$) with compact sizes (log~\reff~($''$) $\lesssim-0.7$), while a positive trend emerges for faint and extended hosts. This suggests that when the SNR of the host is high, the effect of increasing $n$, which controls light concentration, tends to reduce the inferred size. Conversely, for hosts with lower SNR, increasing $n$ encourages the \sersic\ model to extend further into the outer regions, resulting in a larger inferred size. Overall, we emphasize that these changes in host property inferences remain relatively small and fall within our quoted errors. This consistency indicates that our measurements are robust against variations in the \sersic\ index. In the next subsection, we perform further simulation tests to confirm that there are small systematic biases in host galaxy photometry measurements when fixing the \sersic\ $n$ to its true value.

Despite being based on only two-band photometry, our SED fitting yields reliable results for the inference of host stellar mass. The combination of F150W and F356W filters effectively covers the 4000~\AA\ break at $z\sim6$, which is crucial for constraining the host stellar template and estimating stellar mass. Additionally, at high redshift, the age of the universe is relatively small, placing a tight upper limit on the age of galaxies.

\subsection{Image Simulation Tests on Host Galaxy Decomposition}
\label{sec:simulation}

\begin{figure*}
    \centering
    \includegraphics[trim = 80mm 0mm 35mm 0mm, clip, width=1\linewidth]{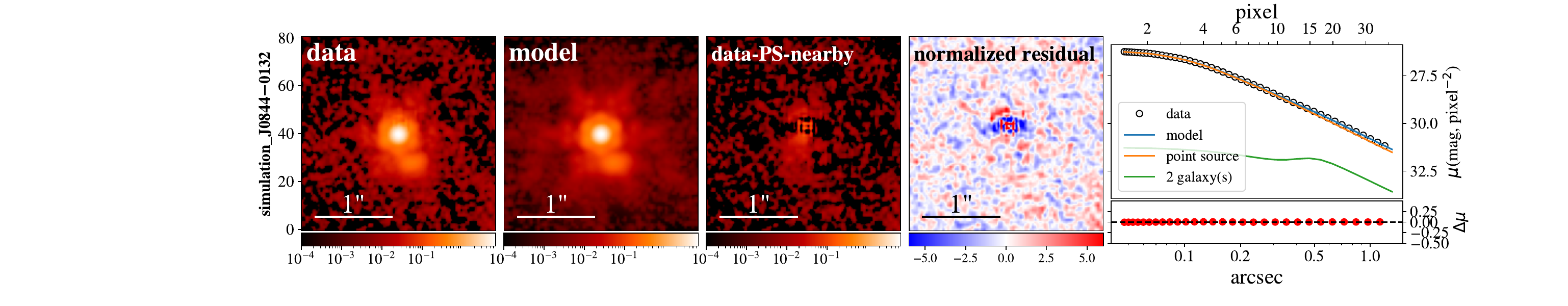}
    \includegraphics[trim = 80mm 0mm 35mm 0mm, clip, width=1\linewidth]{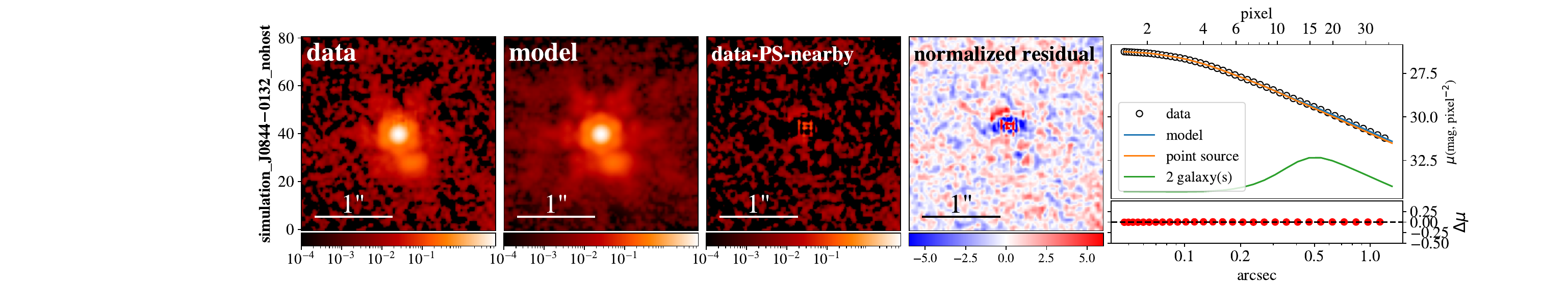}
    \caption{Example of a simulation for a challenging case: J0844$-$0132. (Top:) The mock image is generated by injecting the best-fit quasar, a host galaxy model and a nearby galaxy model into a realistic JWST background, incorporating observational noise and PSF mismatch. In this realization, the recovered host magnitude differs from the input by $-0.2$ mag. The label ``2 galaxy(s)" indicates that we applied two galaxies in the fitting, i.e., the host galaxy and the nearby galaxy. (Bottom:) Same as above, but without injecting a host galaxy component. In this scenario, the host signal in the third panel is no longer present.
    }
    \label{fig:sim_onecase}
\end{figure*}

\begin{figure*}
\centering
\begin{tabular}{c c}
{\includegraphics[trim = 0mm 0mm 0mm 0mm, clip, height=0.45\textwidth]{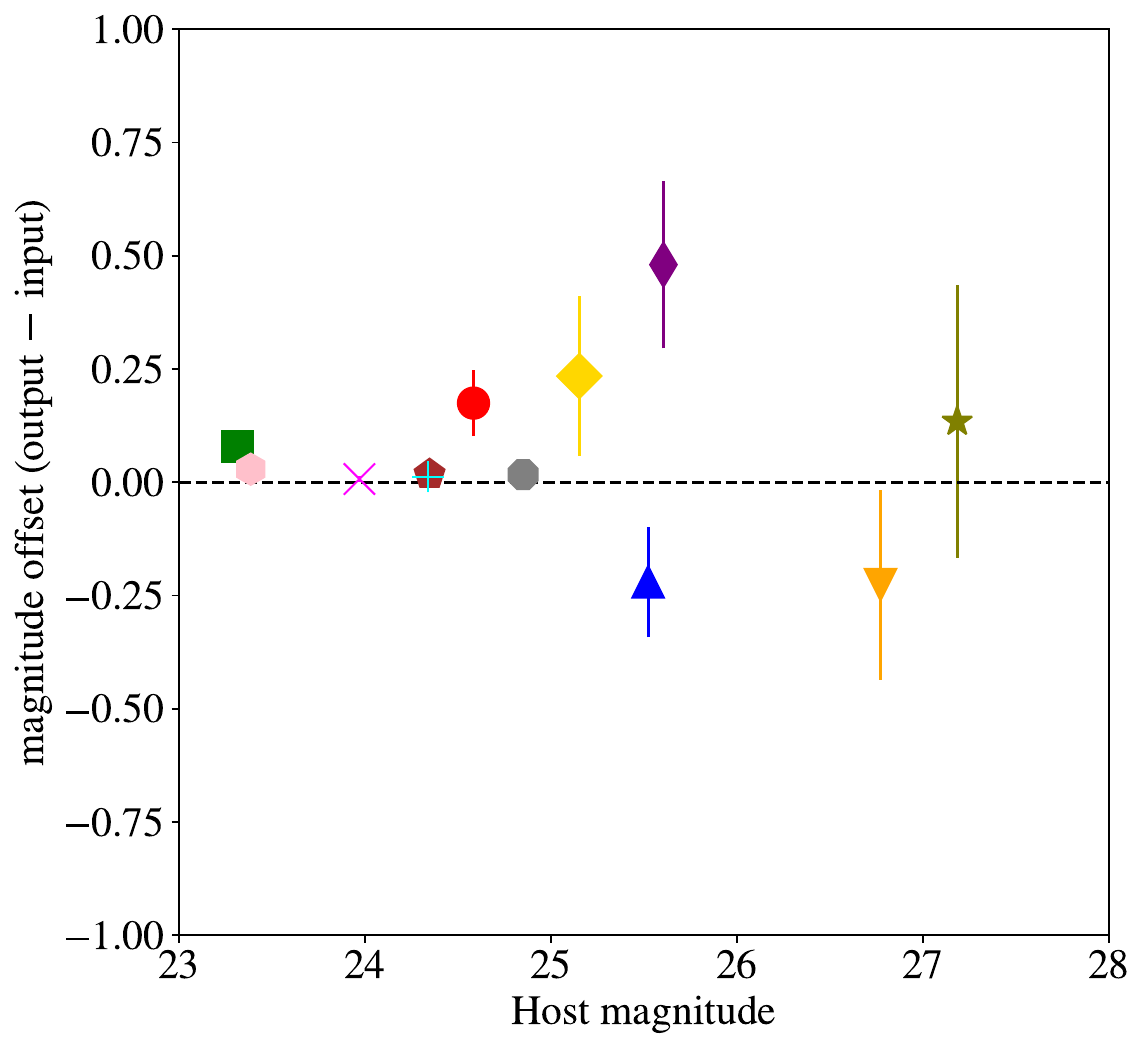}}&
{\includegraphics[trim = 10mm 0mm 0mm 0mm, clip, height=0.45\textwidth]{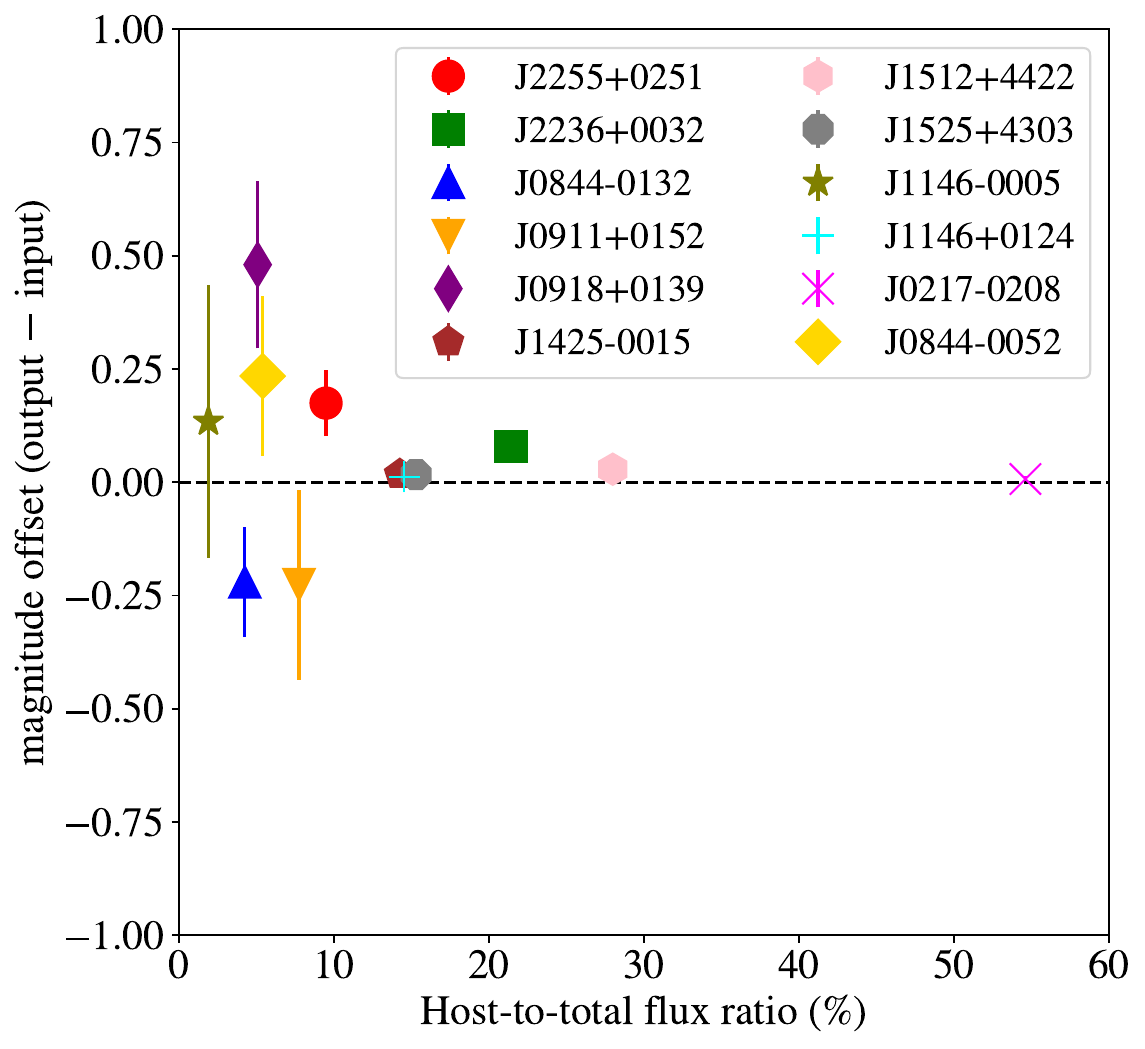}}\\
{\includegraphics[trim = 0mm 0mm 0mm 0mm, clip, height=0.45\textwidth]{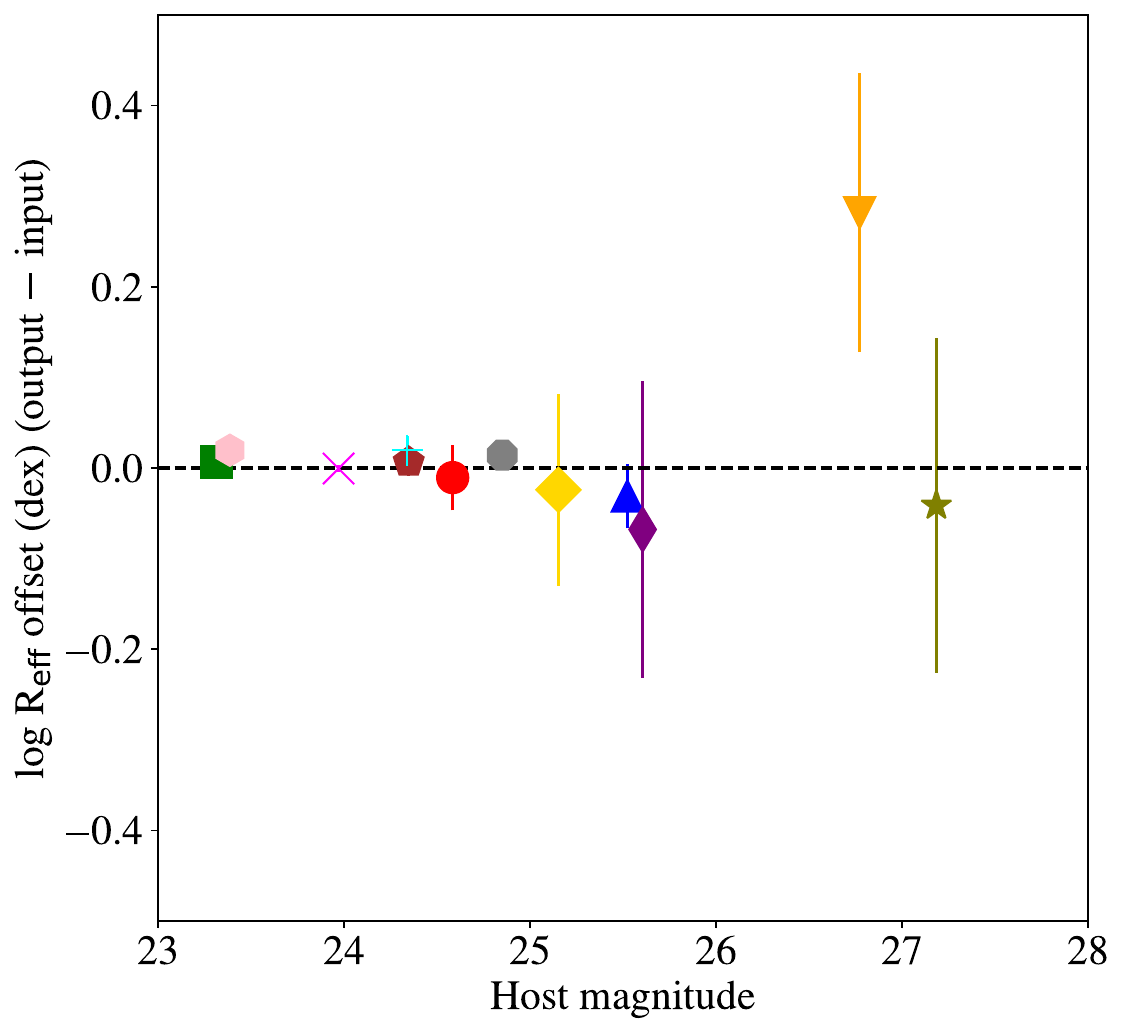}}&
{\includegraphics[trim = 10mm 0mm 0mm 0mm, clip, height=0.45\textwidth]{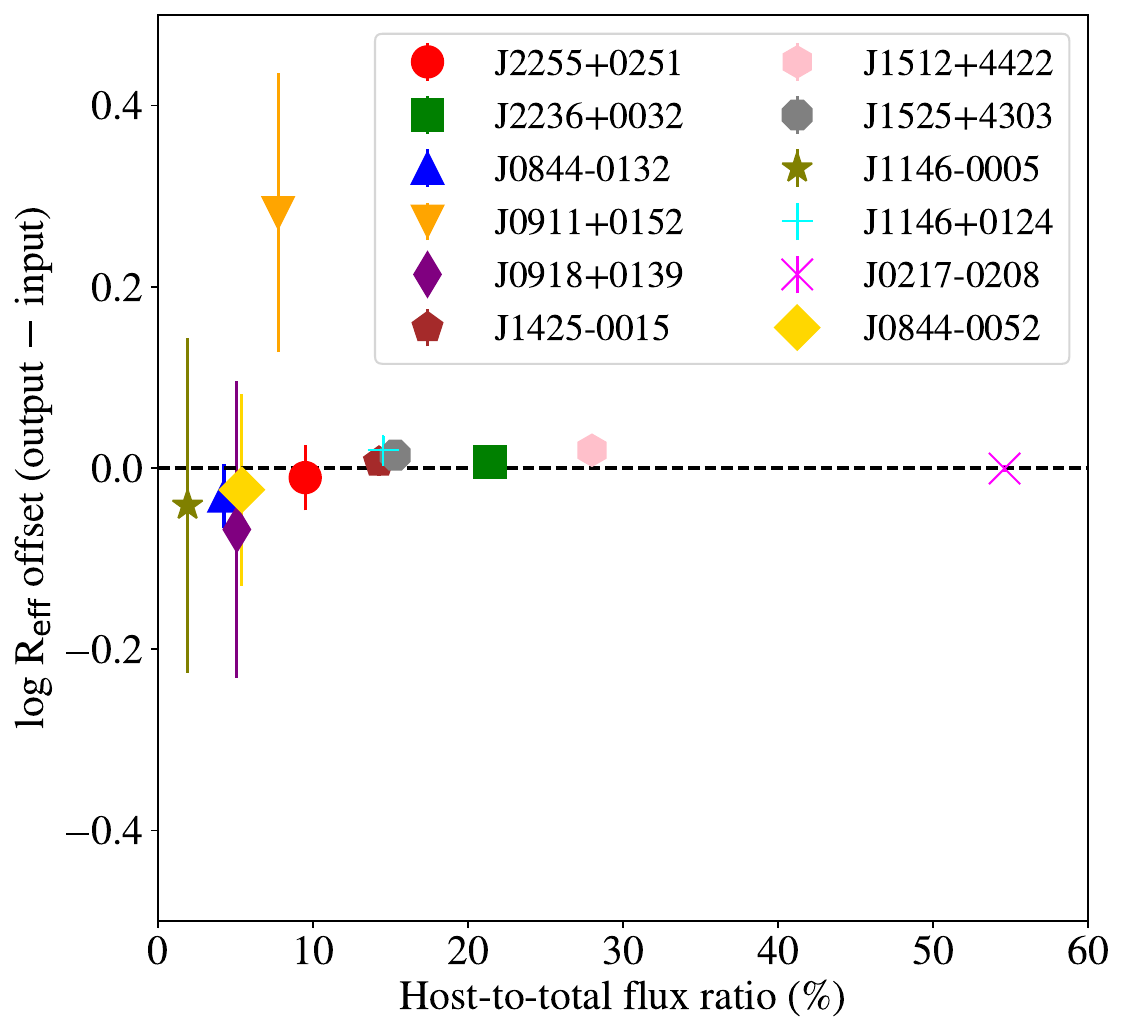}}\\
\end{tabular}
\caption{\label{fig:sim_result} 
\noindent
This figure presents the results of our simulations assessing the recovery of host galaxy properties for all twelve quasar targets in F356W. The top panel shows the recovered host magnitudes minus the truth, while the bottom panel displays the recovered effective radii minus the truth, both plotted as a function of input host magnitude and host-to-total flux ratio. Our results demonstrate that the dispersion in inferred host magnitude increases for fainter hosts (those with magnitudes $>$25) and for systems with host-to-total flux ratios below $10\%$. Nevertheless, the systematic biases in both magnitude and size remain small ($<$0.5 mag, or 0.2 dex), even in cases of non-significant detection such as J1146$-$0005. Additionally, our simulations reproduce the overestimation of host size observed for J0911+0152. Overall, these results validate the robustness of our image decomposition approach and confirm that our methodology yields reliable host property measurements.
}
\end{figure*}

We perform a comprehensive simulation test to evaluate the robustness of our host galaxy measurements. While similar simulations were conducted in~\cite{Ding2023}, we provide more detailed information here. Our simulation tests aim to incorporate the effects of observational noise and PSF mismatch, which are critical factors in high-redshift galaxy studies. To achieve this, we utilize each specific fitting result (adopted from the 40 configurations) and generate mock images that include the inferred light profiles from the quasar, host galaxy, and nearby objects if they exist. These mock images serve as our ``ground truth" based on the best-fit parameters, and we test if our fitting approach could recover them. Specifically, our simulation process involves the following steps:

First, for each target with a specific fitting configuration, we adopt the corresponding inferred host \sersic\ parameters and generate the corresponding \sersic\ profiles, along with the quasar light model, which represents the true host galaxy and quasar emission. We add Poisson noise to these mock images based on the observed exposure time. Next, we insert these mock images into conditions mimicking actual JWST observations, specifically by embedding them into residual images (data minus model). This step ensures that our simulations incorporate realistic background noise and PSF mismatches, closely replicating the actual data environment.

We then process these simulated images through our standard image decomposition pipeline, identical to the approach described in Section~\ref{sec:method}. 
Because our fitting combination strategy effectively reduces systematic uncertainties related to the unknown \sersic\ index, we fix the \sersic\ index to the true value in the simulations to simplify the analysis. We repeat this process for each target, applying it to the top five PSF models to account for PSF variability.

Finally, we compare the inferred host properties from these simulations and combine the comparison results across the top five PSF models to assess their general consistency with the known ``true" input parameters. This process is repeated for all 12 targets, enabling us to evaluate whether our measurements of host magnitude and size are robust within the systematic uncertainties reported in our results. We present one example of our simulation for target J0844$-$0132 in Figure~\ref{fig:sim_onecase} (top).

A summary of our simulation results, demonstrating the accuracy and robustness of the recovered host properties in F356W, is presented in Figure~\ref{fig:sim_result}. Our simulations indicate that systematic biases in host magnitude and size measurements are well within the reported uncertainties (see Table~\ref{tab:F356W_result}), with typical offsets of $<0.30$~mag and $<0.1$~dex, respectively, for both F356W and F150W. The only exception is J0918+0139 in F356W, which exhibits a larger bias of $\sim$0.45 mag (corresponding to an underestimate of $\sim$0.2 dex in stellar mass). Overall, these results validate that our combining strategy (Section~\ref{weighting_stra}) effectively quantifies uncertainties arising from PSF variations and fitting configurations, ensuring robust estimates of host properties.

We also conducted control simulations where we created mock images containing only quasar light, with no host galaxy included. In these cases, our fitting results never simultaneously met the thresholds of a host-to-total flux ratio $>1\%$, host SNR $>0.5$, and $\Delta$BIC$>10$ across all fitting configurations. This confirms that our detection criteria ($>3\%$ flux ratio, SNR $\geq2$ and $\Delta$BIC$>50$) as defined in Section~\ref{sec:detect_crit} are effective at reliably distinguishing true host galaxy detections from false positives caused by fitting artifacts or noise. An example of such a control simulation for the target J0844-0132, without a host galaxy added, is shown in Figure 9 (bottom).


Recent simulation-based work by~\cite{Berger2025arXiv250612130B} provides an independent verification of our host galaxy measurements. Using the hydrodynamical simulation, they perform extensive mock JWST observations and point source removal to statistically quantify potential biases in measured stellar masses of high-redshift quasar hosts caused by quasar subtraction. Their results confirm that the methodologies employed in our study reliably recover intrinsic host galaxy magnitudes and stellar masses within modest uncertainties, thereby reinforcing the robustness of our host decomposition and photometric analysis techniques.

\section{Summary} \label{sec:sum}
We study a sample of twelve moderate-luminosity quasars ($M_{1450}>-24$) observed with JWST/NIRCam in the F150W and F356W filters. We characterize their host galaxies through a state-of-the-art 2D image decomposition technique. 
To address systematic uncertainties in measuring host properties, such as host flux and effective radius, we adopt a weighting algorithm based on different fitting settings (Section~\ref{weighting_stra}). 
We establish three key criteria that must be satisfied to classify the quasar host galaxy as a significant detection (Section~\ref{sec:detect_crit}).
Additionally, we employ two-band spectral energy distribution (SED) fitting to derive the stellar masses of the host galaxies (Section~\ref{subsec:sed}). Our results include the size--mass relation, stellar mass density, and local environment of our quasar hosts, which we compare with that of a normal (i.e., non-quasar) galaxy sample observed by the COSMOS-Web survey and massive quiescent galaxies at $z\gtrsim4$ in the literature.

We summarize the key findings of this work as follows:

\begin{itemize}
\item We successfully resolve and detect host galaxies for 11/12 quasars in F356W and 7/12 in F150W (see Figure~\ref{fig:F356Wfit} and Figure~\ref{fig:F150Wfit}), demonstrating the capability of JWST/NIRCam to separate host emission from bright quasar cores at $z>6$.

\item We find that the host galaxies span a wide range of stellar masses (log~M$_*$/M$_{\odot} = 9.5\text{--}11.0$, Table~\ref{tab:table4}) and sizes (0.5$-$3 kpc, at the rest-frame optical wavelength by F356W, Table~\ref{tab:F356W_result}), with compact morphologies prevalent among massive systems.

\item The size--mass relation of quasar hosts is generally consistent with that of non-quasar star-forming galaxies in COSMOS-Web at $z\sim6$ (Figure~\ref{fig:size-mass}, left), suggesting shared structural evolution pathways during the epoch of reionization.

\item The most massive hosts of our sample (i.e., those with log~M$_{\rm \odot}>$10.5) and the two post-starbursts (J2236+0032 and J1512+4422) exhibit ultra-compact sizes and elevated stellar mass densities, similar to those observed in $z\sim4-5$ quiescent galaxies in COSMOS-web. This supports models where gas inflows drive central starbursts followed by rapid quenching.

\item Half of our quasar hosts have nearby companions (within a projected separation of 2\farcs1 ($\sim12$kpc) in the F356W images); this fraction is consistent with that observed for non-quasar star-forming galaxies in COSMOS-Web at $z\sim6$ at a similar galaxy stellar mass. However, the decomposed quasar host light does not show prominent merger signatures, suggesting that minor interactions or secular processes, such as disk instabilities, play a dominant role in fueling AGN at $z>6$.


\item The eight-band observations of quasar host galaxy J2236+0032 reveal consistent morphology across different wavelengths, see Table~\ref{table:J2236_result} and Figure E1a in~\cite{Onoue2024}. Furthermore, the agreement between imaging-based decomposition (using NIRCam) and spectral decomposition~\citep[by][using NIRSpec]{Onoue2024} validates the robustness of our image decomposition framework, ensuring accurate characterization of host galaxy properties.
\end{itemize}

Our findings support galaxy compaction models, which propose that gas-rich disks undergo turbulent inflows driven by mergers, instabilities, or AGN feedback, triggering centrally concentrated starbursts that build dense cores. The most massive quasar hosts in our sample (log~M$_{*}$/M$_{\rm \odot}>$10.5) exhibit ultra-compact sizes and elevated stellar mass densities, resembling $z\sim4-5$ quiescent galaxies, suggesting that compaction-driven starbursts, followed by rapid quenching, may already be shaping the structural evolution of massive galaxies during the reionization epoch. The two post-starburst quasar hosts (J2236+0032 and J1512+4422) further align with this scenario, as their compact morphologies and high mass densities are consistent with the properties of quiescent galaxies (Figure~\ref{fig:size-mass}, right), indicating a potential transition toward quiescence.

To confirm this compaction-driven evolutionary scenario, it is essential to gather more measurements of quasar hosts across a broader redshift range ($2<z<6$). Moderate-luminosity quasars ($M_{1450}>-24$ mag) are critical for this effort: their lower AGN-host contrast enables a high success rate of host galaxy detections, while their prevalence minimizes biases inherent in luminous systems. By extending observations to intermediate redshifts (i.e., $z\sim3-5$), we can clarify the roles of secular processes (e.g., disk instabilities) versus mergers in fueling AGN activity and shaping host morphologies. Such an expanded dataset would also allow us to directly investigate evolutionary processes, such as minor mergers and structural instabilities, that drive changes in galaxy sizes and densities. 

\begin{acknowledgments}
We sincerely thank  Xiaohui Fan and Shenli Tang for valuable discussions and insightful suggestions.

This work is based on observations made with the NASA/ESA/CSA James Webb Space Telescope. The data were obtained from the Mikulski Archive for Space Telescopes at the Space Telescope Science Institute, which is operated by the Association of Universities for Research in Astronomy, Inc., under NASA contract NAS 5-03127 for JWST. These observations are associated with programs GO \#1967, GO \#3859 and GO \#1727. Support for these programs was provided by NASA through a grant from the Space Telescope Science Institute, which is operated by the Association of Universities for Research in Astronomy, Inc., under NASA contract NAS 5-03127. This work was supported by World Premier International Research Center Initiative (WPI), MEXT, Japan. This work used computing resources at Kavli IPMU.
All the  {\it JWST} data used in this paper can be found in MAST: \dataset[10.17909/hqaf-an74]{http://dx.doi.org/10.17909/hqaf-an74}.

Support for this work was provided by NASA through grant JWST-GO-01727 awarded by the Space Telescope Science Institute, which is operated by the Association of Universities for Research in Astronomy, Inc., under NASA contract NAS 5-26555.
MO is supported by the Japan Society for the Promotion of Science (JSPS) KAKENHI grant No. G24K22894.
YM is supported by the Japan Society for the Promotion of Science (JSPS) KAKENHI Grant No. 21H04494.
SEIB is supported by the Deutsche Forschungsgemeinschaft (DFG) under Emmy Noether grant number BO 5771/1-1.
JS is supported by JSPS KAKENHI (JP22H01262) and the World Premier International Research Center Initiative (WPI), MEXT, Japan.
KI acknowledges support from the National Natural Science Foundation of China (12073003, 11721303, 11991052).
AL acknowledges support from PRIN MUR 2022 - Project “2022935STW"
JTS is supported by the Deutsche Forschungsgemeinschaft (DFG, German Research Foundation) - Project number 518006966.
M.V. gratefully acknowledges financial support from the Independent Research Fund Denmark via grant numbers DFF 8021-00130 and  3103-00146. KI acknowledges support under the grant PID2022-136827NB-C44 provided by MCIN/AEI/10.13039/501100011033 / FEDER, UE.
FW acknowledges support from NSF award AST-2513040.
\end{acknowledgments}

%

\vspace{5mm}
\facilities{JWST/NIRCam.}

\software{\galight~\citep{Ding2020}, {\sc lenstronomy}~\citep{Birrer2018, Birrer2021}, {\sc gsf}~\citep{Morishita2019},  
{\sc astropy}~\citep{2013A&A...558A..33A,2018AJ....156..123A}, 
{\sc photutils}~\citep{Photutils2016ascl.soft09011B}
          }

\bibliographystyle{aasjournal}



\end{document}